\documentclass[12pt]{article}
\usepackage{graphicx}

\newcommand{\COMMENTO}[1]{}
\newcommand{\ddh}{{\hat d}}

\newcommand{\bb}{{\cal B}}
\newcommand{\cc}{{\cal C}}
\newcommand{\dd}{{\cal D}}
\newcommand{\ee}{{\cal E}}

\newcommand{\eq}[1]{eq.~(\ref{#1})}

\def\reali{{\hbox{l\kern-.5mm R}}}
\def\complessi{{\hbox{l\kern-1.9mm C}}}
\def\Hplus{{\hbox{l\kern-1.9mm H}}^+}

\def\interi{{\mathchoice
 {\hbox{Z\kern-1.5mm Z}}
 {\hbox{Z\kern-1.5mm Z}}
 {\hbox{{Z\kern-1.2mm Z}}}
 {\hbox{{Z\kern-1.2mm Z}}}  }}
\def\Z{\interi}
\def\uno{{\hbox{1\kern-0.3em 1}}}

\newcommand{\sect}[1]{\setcounter{equation}{0}\section{#1}}

\newcommand{\YZp}{{ -y_0}}
\newcommand{\YZ}{{}}
\newcommand{\YZS}{{}}
\newcommand{\EYZi}{}

\renewcommand{\thefootnote}{\fnsymbol{footnote}}

\begin{document}
\begin{titlepage}
\rightline{DFTT 19/03}
\rightline{\hfill September 2003}

\vskip 1.2cm

\centerline{\Large \bf 
Multi-branes boundary states with open string interactions
}

\vskip 1.2cm

\centerline{\bf Igor Pesando\footnote{e-mail:
ipesando@to.infn.it}}
\centerline{\sl Dipartimento di Fisica Teorica, Universit\`a di
Torino}
\centerline{\sl Via P.Giuria 1, I-10125 Torino, Italy}
\centerline{\sl and I.N.F.N., Sezione di Torino}

\begin{abstract}
We derive boundary states which describe 
configurations of multiple parallel branes with arbitrary open string
states interactions in bosonic string theory. 
This is obtained by a careful discussion of the
factorization of open/closed string states amplitudes taking care of
cycles needed by ensuring vertices commutativity: in particular the
discussion reveals that already at the tree level open string knows of
the existence of closed string.
\end{abstract}

\end{titlepage}

\newpage
\COMMENTO{      
}
\renewcommand{\thefootnote}{\arabic{footnote}}
\setcounter{footnote}{0}
\setcounter{page}{1}

%%%%%%%%%%%%%%%%%%%%%%%%%%%%%%%%%%%%%%%%%%%%%%%%%%%%%%%%%%%%%%%%%%%%%%%%%%%%%%
%%%%%%%%%%%%%%%%%%%%%%%%%%%%%%%%%%%%%%%%%%%%%%%%%%%%%%%%%%%%%%%%%%%%%%%%%%%%%%
%%%%%%%%%%%%%%%%%%%%%%%%%%%%%%%%%%%%%%%%%%%%%%%%%%%%%%%%%%%%%%%%%%%%%%%%%%%%%%

\sect{Introduction and motivations.}
\label{intro}
Since the discovery of the non perturbative role played by strings
with Dirichlet boundary conditions \cite{Polchinski:1995mt} the
boundary state formalism, first used in \cite{bounstate}, has
been further developed.
It has proven to be useful in a number of situations such as, for
example, in reading which supergravity fields are switched on in
presence of branes \cite{DiVecchia:1997pr,Bertolini:2000dk} 
or in analyzing the rolling of the tachyon (see for example \cite{Sen:2002nu}). 

All the boundary states used were trivially a superposition
of boundary states and they did not describe any open string
interactions (see however app. A of the first of \cite{bounstate}):
this can be rephrased technically by saying that the boundary state 
of $N$ branes was taken to be
equal to $N|B\rangle$ where $|B\rangle$ is the boundary state of a
single brane. 
This leaves the doubt whether it is possible to describe with the
boundary state formalism open string interactions or a non trivial
superposition of branes.

The aim of this paper is to show that it is actually simple to
describe open/open and open/closed string interactions 
within the closed string formalism
and that it is also possible to describe with a unique boundary state
a set of several parallel branes interacting among themselves in the open string channel:
an example of the amplitudes described by the boundaries we consider
can be seen in fig.s (\ref{figure:3open-1closed}, \ref{figure:disk-3open-1closed}).
In doing so we notice how the open string knows about the closed
string already at the tree level.
We will show this starting from mixed closed/open string amplitudes
and then factorizing  them in the closed channel in order to get
boundary states which describe open string interactions expressed using
closed string operators: the resulting boundary state (see
eq. (\ref{B_alpha})) is for equal parallel
branes in generic position, i.e. when the branes are not superimposed
and the low energy gauge group is a
product of one $U(1)$ for each brane. It is nevertheless easy to get
the non generic situation with enhanced gauge group (see eq. (\ref{B_alpha_CP})).

Since it is possible to describe open string interactions with closed
string formalism it is natural to ask whether it is possible to do the
opposite, i.e. to describe pure closed string interactions within open string
formalism. 
\COMMENTO{ 
??????
The formal answer is yes since given any closed string amplitudes on a
disk ${\cal A}(\{\beta_L,\beta_R\})(f_j)$
where the underlying open string is coupled to a constant
electromagnetic background in all directions $F_{2j, 2j+1}=f_j$ with
$j=0,\dots,\frac{D-1}{2}$ it is possible to formally compute the associated
pure closed string amplitude  by computing 
$$\prod_j \oint_{|f_j -i|<2} \frac{d f_j}{2\pi i}
\frac{f_j}{(1+f_j^2)^{3/2}} 
{\cal   A}(\{\beta_L,\beta_R\})(f_j) =
{\cal A}_{\mbox{pure closed string}}(\{\beta_L,\beta_R\})$$
this is easy to show in the boundary formalism where 
\begin{eqnarray*}
&\prod_j \oint_{|f_j -i|<2} \frac{d f_j}{2\pi i}
\frac{f_j}{(1+f_j^2)^{3/2}}|B(f)\rangle
\,\,\propto&
\\
&
\prod_j \oint_{|f_j-i|<2} \frac{d f_j}{2\pi i}
\frac{f_j}{(1+f_j^2)}
e^{-\frac{1}{1+f_j^2}
\sum_n (a^\dagger_{2 j},a^\dagger_{2 j+1})
\left(\begin{array}{cc}
\scriptstyle
1-f_j^2 & \scriptstyle f_j\\
\scriptstyle -f_j & \scriptstyle 1-f_j^2
\end{array} \right)
\left( \begin{array}{c}
\scriptstyle \tilde a^\dagger_{2 j} \\ 
\scriptstyle \tilde a^\dagger_{2 j+1}
\end{array} \right)
}
|0,\tilde 0>
&
\end{eqnarray*}
}
An almost positive answer is given in \cite{Gaiotto:2003rm}
where closed string amplitudes (with a special state insertions) are
obtained from an ordered sequence of D-branes located at imaginary time.

The paper is organized as follows.
In section \ref{par1} we fix our notations 
(see also appendix \ref{conventions}) by reviewing how to compute mixed
open/closed string amplitudes in open string formalism and
we comment also on how Chan-Paton factors emerge naturally when branes
are moved from a generic position where they are parallel but non
superimposed to a superimposed configuration. 
We then write the vertices which describe the emission of closed string
states using only open string fields and vertices for 
the emission of open strings hanging 
between two parallel but separated branes  up to cocycles. 
The cocycles are discussed in section \ref{par2} for both open and
closed string formalism: they are necessary to ensure the vertices 
commutativity in both closed and open formalism and allow a consistent
factorization of amplitudes. 
We start computing the cocycles in closed string formalism since the
vertices for the emission of closed string states in open string
formalism must reproduce the same OPEs as in the closed string sector.
It is worth stressing that in order to achieve the full consistency of the
theory is necessary not only to introduce cocycles but also to let $y_0$
(which is the position of the $\sigma=0$ boundary of the string  
when we T-dualize to a configuration where all space directions have 
 DD boundary conditions or the Wilson lines on the brane at $\sigma=0$
 when all directions are NN) 
change after the emission of an open string.
This change is also necessary and felt by closed string vertices when all
boundary conditions are NN.
Moreover it is interesting to notice that cocycles
are not invariant under T-duality and therefore amplitudes are
determined up to an overall sign. 

In section \ref{par3} using the classical techniques
we perform the factorization of the open string formalism amplitudes
into the closed string channel and we obtain the explicit form of
boundary states describing a bunch of parallel branes interacting in
the open channel. These boundary states can be used, for example, to recover
mixed string amplitudes of one closed string state with many open
string states simply by performing the product with the
closed string state or to obtain one loop amplitudes by computing
boundary-boundary interaction. 

Finally in section \ref{par-conc} we draw our conclusions.

%%%%%%%%%%%%%%%%%%%%%%%%%%%%%%%%%%%%%%%%%%%%%%%%%%%%%%%%%%%%%%%%%%%%%%%%%%%%%%
%%%%%%%%%%%%%%%%%%%%%%%%%%%%%%%%%%%%%%%%%%%%%%%%%%%%%%%%%%%%%%%%%%%%%%%%%%%%%%
%%%%%%%%%%%%%%%%%%%%%%%%%%%%%%%%%%%%%%%%%%%%%%%%%%%%%%%%%%%%%%%%%%%%%%%%%%%%%%

\sect{Review of mixed closed and open string amplitudes %and some
%  comments on Chan-Paton factors
.}
\label{par1}

Interactions among closed and open strings were extensively
studied already in the early days of string theory
\cite{old,Ademetal}. In particular, in Ref.~\cite{Ademetal}
Ademollo et al. constructed
vertex operators for the emission of a closed string out of an open
string, and computed the scattering amplitudes among $N_c$ closed
and $N_o$ open strings at tree level. The topology of the
string world-sheet corresponding to these amplitudes is
that of a disk emitting $N_o$ open strings
from its boundary and $N_c$ closed strings from its interior.
As customary in those days, only Neumann boundary conditions
were imposed on the disk, and no target-space compactification was
considered. 
This formalism was extended to the case of mixed Neumann-Neumann and
Dirichlet-Dirichlet 
boundary conditions in a compactified target space in Ref.~
\cite{DiVecchia:1997pr}.

We find now useful to recall here these results also in order to fix
our notations.

We consider a bosonic string described by $X^\mu$ ($\mu=0,\dots,D-1$)
propagating in a partially compactified space-time with 
toroidal compact directions $X^a\equiv X^a+2\pi R^a$
($a=\ddh,\dots,D-1$), to simplify the notation we will generically set 
all $R^a=R$.
Both compact and non compact directions can have either NN  or DD
boundary conditions.
%By a conformal transformation a disk can be mapped onto the upper half
%complex $z$-plane ($z={\rm e}^{\tau+i\sigma}$ with $\tau$ and
%$\sigma$ ($0\le\sigma\le\pi$)
%being the Wick rotated timelike and spatial coordinates of the world
%sheet) and
%its boundary onto the real axis ($x=\pm{\rm e}^{\tau}$). 

We write the open string coordinates as a function of left and right 
components $X_L(z)$, $X_R(\bar z)$ as
\begin{equation}
\label{split}
X(\tau,\sigma)={1\over 2}\left[X_L(z)+X_R(\bar z)\right]~~~,
\end{equation}
independently of the boundary conditions.
In the previous expression \eq{split} $X_{L,R}$ are defined as
\begin{equation}
\label{splitb}
\left\{
\begin{array}{ccc}
X^{\mu}_{L}(z)  & = & 
q^{\mu}+ y^\mu_0 
  -{\rm i}\, \sqrt{2\alpha'}\, a_0^{\mu}{}\ln z +
  {\rm i}\,\sqrt{2\alpha'}\,
  \sum_{n\not =0} {sgn(n)\over {\sqrt {|n|}}} a_n^{\mu}z^{-n}
\\
X^{\mu}_{R}(z)  & = & 
S^\mu_\nu \left(
  q^{\nu}- y^\nu_0 
  -{\rm i}\, \sqrt{2\alpha'}\, a_0^{\nu}{}\ln {\bar z} +
  {\rm i}\,\sqrt{2\alpha'}\,
  \sum_{n\not =0} {sgn(n)\over {\sqrt {|n|}}} a_n^{\nu}{\bar z}^{-n}
  \right)
\end{array}
\right.
\label{XL-XR}
\end{equation}
where 
$-\pi<arg(z)\le + \pi$, $\log({\bar z})=\overline{\log(z)}$,
 $y_0$ is an operator which commute with everything and can be traded
with a constant (see later for more) 
and
in the second equation we have introduced the diagonal matrix 
\begin{equation}
||S^\mu_\nu||=diag(\pm 1,\pm 1,\dots,\pm1)
\end{equation} 
in which all the signs $+(-)$ can be chosen independently.
Each sign is fixed to $+(-)$ 
when the corresponding coordinate has NN(DD) boundary condition,
explicitly if we
consider a NN direction we take $S=1$ and we have
($z=e^{i(\tau+\sigma)}=e^{\tau_E+i\sigma}$ where $\tau_E$ is the Wick
rotated time)
\begin{equation}
X(\tau,\sigma)= q_0 + 2\alpha' p \tau + oscillators ,
\end{equation}
while for a DD direction we choose $S=-1$ and we find
\begin{equation}
X(\tau,\sigma)= y_0 + 2\alpha' p \sigma + oscillators .
\end{equation}
From these expressions
it is clear that $y_0$ plays the role of the boundary value in case of
a direction with DD boundary condition,
i.e. for example in compact directions $X^a_{DD}|_{\sigma=0}=y_0^a$ and
$X_{DD}^a|_{\sigma=\pi}=y_\pi^a=y_0^a+2 \pi (w^a+\theta^a) R'$
where $R'=\frac{\alpha'}{R}$ is the ``dual'' radius, 
the operator $\alpha' p^a$ has spectrum
$ \alpha' p^a=  \alpha' \frac{w^a+\theta^a}{R}=(w^a+\theta^a) R'$
 and $0\le \theta^a <1$ is associated
with  the difference of position of two branes or, with NN boundary
conditions, with
the difference of Wilson lines which can be eventually turned on
on the $\sigma=0$ and $\sigma=\pi$ branes.% (see eq. (\ref{momentum}) ).

It is also very important to point out that the zero mode part of the
expansion in the $DD$ sector 
$X_{DD(zero~ modes)}=y_0
-{\rm i}\, \sqrt{2\alpha'}\, a_0^{\mu}{}\ln \sqrt{z\over \bar z} $
is an operator 
(where $a_0= \sqrt{2\alpha'} p$ 
%has a spectrum given by $\frac{(w+\theta) R'}{\alpha'}$ with $w\in\interi$
) 
and   not a c-number as one would deduce from the
canonical quantization, exactly as it happens for the winding number
for a closed string propagating on a torus.
This fact is fundamental in the construction of
{\sl massive} $W^\pm$  emission vertices \cite{Pesando:1999hm} and in the
following construction of the cocycles.
It is noteworthy to stress once again that $y_0$ is an operator too but it
commutes with everything and therefore it can be traded with its
eigenvalues and treated as a c-number: the same happens with
the usual Chan-Paton factors and this is not accidental since $y_0$ is a
kind of Chan-Paton factor as we discuss below.

The emission from the disk of an open string state with
internal quantum numbers $\alpha$ is described 
by a vertex operator ${V}_{\alpha}(X_L(x);k)$ up to a possible
cocycle to be discussed later.
The vector $k_\mu=(k_\alpha ,k_a)$ ($\alpha=0,\dots,d-1)$) 
 must be interpreted as 
$k=\left\{
\begin{array}{lc} 
momentum &  \textrm{NN bc}\\
\frac{distance}{2\pi\alpha'} & \textrm{DD bc}
\end{array} 
\right.$,
i.e. $momentum$ in directions with NN boundary conditions 
and as ${distance}/(2\pi\alpha')$ in the ones with DD boundary 
conditions\footnote{
In presence of $DD$ direction this
requires a little explanation, see \cite{Pesando:1999hm} and also our further
discussion in section 3.2).
The string endpoints can be attached on two different branes at a distance $\delta$ 
in such a DD direction. 
This changes the mass shell condition due to the energy coming from
the tension of the stretched string while the vertex still must have conformal
dimension one, hence we must explicitly act on the $SL(2,\reali)$
vacuum with an operator to get this distance.
}.
In particular for compact NN directions we have
$k^a=\frac{k^a+\theta^a}{R}$  where $n^a \in \interi$.

The emission of a closed string state 
with momentum in non compact direction $k^\alpha=2k^\alpha_L=2k^\alpha_R$, 
momentum 
${k_L^a} = {1\over 2}\left({n^a\over R}+ {w^a R\over \alpha'}\right)$,
${k_R^a}  =  {1\over 2}\left({n^a\over R} - {w^a R\over \alpha'}\right)$
in compact direction
 and left and right quantum numbers $\beta_L$ and $\beta_R$, is
described by a vertex operator ${W}_{\beta_L,\beta_R}(z,{\bar z};k_L,k_R)$:
the presence of a boundary on the world sheet imposes
a relation between the left and right
parts of ${W}_{\beta_L,\beta_R}$ which are not
independent of each other. 
In fact it is possible to write \cite{Ademetal,DiVecchia:1997pr}
\footnote{
One could wonder why not to use ${ W}'={ V}_{\beta_R} { V}_{\beta_L}$
instead:
the two vertices ${ W}$ and ${ W}'$ differ
in fact by a phase $\exp{(i ~2\alpha' ~\pi k_L\cdot k_R)}$ which is
non trivial in presence of compact directions.
%The answer is that amplitudes on compactified space are determined up
%to a phase (common to all loop) as it will become clearer when
%discussing closed string cocycles.
The answer is that the true vertices need a cocycle and that the two
versions of the vertices must have different cocycles since they both must
reproduce the closed OPEs.
%Nevertheless these cocycles differ in such a way 
%just for the same phase (up to multiple of $\pi$ since amplitudes on
%compactified space are determined up to a sign common to all loop).
}
\begin{equation}
\label{Wertex}
{ W}_{\beta_L,\beta_R}(z,\bar z; k_L,k_R) = 
{ V}_{\beta_L}(X_L(z);k_L)
{}~{V}_{\beta_R}(X_R(\bar z); k_R)~~~,
\end{equation}
again up to a cocycle. %and where $k_L^2-k_R^2\in\interi$. 
We would like to stress that the vertex operator
${ W}$ depends on a single set of oscillators ({\it i.e.} those of
the  open string), and that each factor in \eq{Wertex} is separately
normal ordered.
This is to be contrasted with the
vertex operators describing the emission of a closed string
out of a closed string where there are
two distinct sets of oscillators for the left and right sectors which
only share the zero-mode in the non compact case.

A simple and intuitive reason why it is possible to write closed
 string vertices using open string fields  is that if we
 start  with the free
open string sigma model with whichever boundary condition we can always
couple the graviton as
\begin{equation}
S=\frac{1}{4\pi\alpha'}\int d^2\xi \eta_{\mu\nu}\partial X^\mu
\partial X^\nu 
\rightarrow 
S=\frac{1}{4\pi\alpha'}\int d^2\xi g_{\mu\nu}(X) ~\partial X^\mu
\partial X^\nu
\label{sigma_model}
\end{equation}
and this means that the graviton vertex which follows from the weak
coupling expansion $g_{\mu\nu}(X)=\eta_{\mu\nu}+\kappa h_{\mu\nu}(X)$ 
can be expressed through the open string fields $X$. 
In particular the quantum conformal
properties required for a vertex implies its factorized form as shown
in eq. (\ref{Wertex}) for a generic closed string state.
From the same argument this result applies to all the massless closed string
excitations and it can be extended to massive states by OPE.
It is in this process that cocycles become unavoidable since otherwise
the OPE coefficients in closed and open formalism 
would differ by phases.

% and thus is a primary field of weight one.

\begin{figure}[hbtp]
\begin{minipage}[t]{0.45\linewidth}
\scalebox{0.9}{
\includegraphics[type=eps,ext=.eps,read=.eps,width=0.9\textwidth]{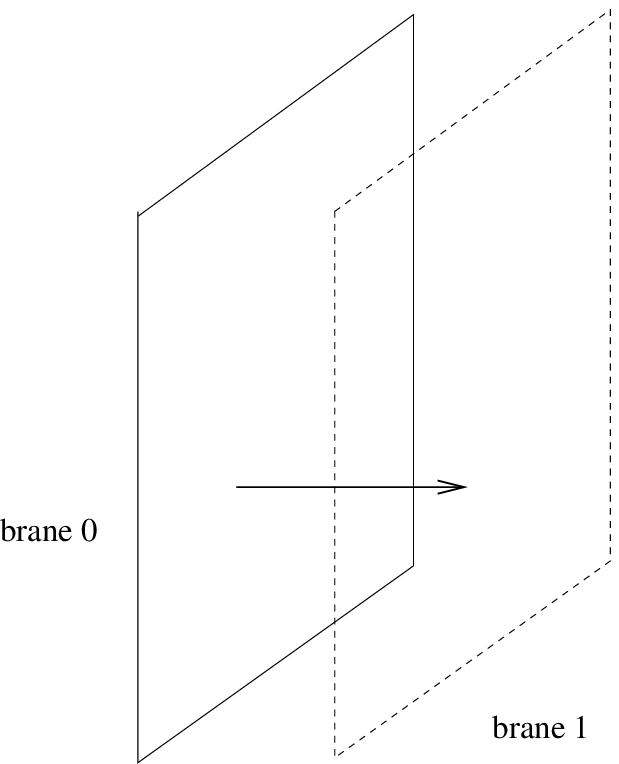}
}
 \caption{As long as brane 0 is at $y_0$ and brane 1 at $y_0+2\pi
   \alpha' p$ with $p\ne 0$
the brane positions are more informative than
Chan-Paton factors and are sufficient to identify the branes.
When $p=0$ the branes are superimposed and Chan-Paton factors are
   needed since all branes have the same position.
  }
\label{figure:2branes}
\end{minipage}
\hskip 1cm
\begin{minipage}[t]{0.45\linewidth}
\scalebox{0.9}{
\includegraphics[type=eps,ext=.eps,read=.eps,width=0.9\textwidth]{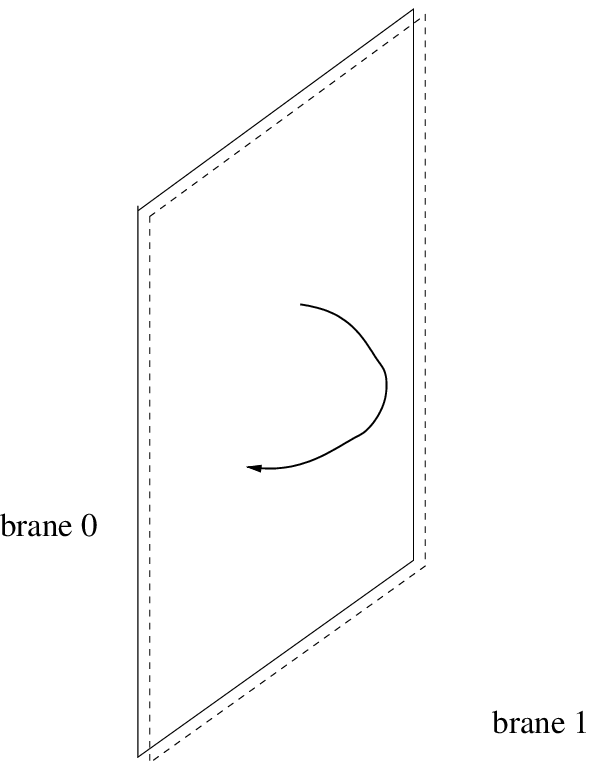}
}
\caption{ When the branes are coincident one must introduce Chan-Paton
since we cannot anymore label the branes by their position.}
\label{figure:2superimposed_branes}
\end{minipage}
\end{figure}

Using the complete operators ${\cal V}$ and ${\cal W}$ (i.e. with the
cocycles included),
the tree-level scattering amplitude among $N_o$ open and $N_c$ closed strings
is given by %($x_{N_o+1}=\infty$)
\begin{eqnarray}
\label{Ademampl}
{\cal A}(N_o, N_c) & = &
{\cal C}_{0(p)}\,{\cal N}_{o(p)}^{N_o} 
\,{{\cal N}_c}^{N_c}
\int {1\over d V_{abc}}~
d x_{N_o}
~\prod_{i=1}^{N_o-1} \left[d x_i\;\theta(x_{i+1}- x_i)\right]
~\prod_{j=1}^{N_c} d^2z_j 
\nonumber\\ 
& & \langle 0 |~{\rm T}\left(\prod_{i=1}^{N_o} {\cal V}_{\alpha_i}(x_i;k_i) 
\,\prod_{j=1}^{N_c} {\cal W}_{\beta_{j L},\beta_{j R}}(z_j, \bar z_j;
k_{L~j}, k_{R~j}) \right)
| 0\rangle~~~,
\end{eqnarray}
where
$d V_{a b c}$ is the volume of the projective group $SL(2,{\bf R})$,
T denotes the time (radial) ordering prescription, and
${\cal C}_{0 (p)}$, ${\cal N}_{o (p)}$ and ${{\cal N}_c}$ are respectively
the normalizations of the disk, of the open and of the closed
vertex operators corresponding to a $Dp$ brane: they are explicitly
given in appendix \ref{conventions}.
They are fixed so that
the low energy limit of the previous amplitudes reproduces the
corresponding field theoretical amplitudes of a theory living in $p+1$
spacetime dimensions with $D_{NN,c}$ compact ones
(see  ref.~\cite{1loop}).
In \eq{Ademampl} the variables $x_i$'s are integrated on the real
axis while the complex variables $z_j$'s are integrated
on the upper half complex plane.
%Because of \eq{Wertex} it is clear that $A(N_o,N_c)$
%is formally similar to a pure open string amplitude with
%$N_o+2 N_c$ external states provided suitable identifications
%of momenta are made. This was also re-proposed in a simplified case in
%\cite{Myers}.
In this expression we have not used  Chan-Paton factors since
we suppose the branes are in a generic position as in fig.
\ref{figure:2branes}, i.e. they are parallel
but do not overlap (hence we are restricting to the case of $Dp$ with $p
< 25$) 
and therefore the gauge group is a product of 
 $U(1)$ factors.
Nevertheless it is possible to introduce the Chan-Paton factors
$\Lambda$ simply by performing the substitution 
${ V}_{\alpha}(x;k) \rightarrow { V}_{\alpha}(x;k) \Lambda$
when it is necessary, i.e. when 
the branes are not anymore in a generic position and they are stacked.

The reason why we did not write the Chan-Paton factors in
eq. (\ref{Ademampl}) is that they are not necessary when branes are in
a generic position.
Let us see why.
A Chan-Paton factor is nothing but a matrix:
%which can be associated to a couple of labels: 
in the simplest setting, i.e. for gauge group
$U(N)$, we can take as the Chan Paton matrices 
$\Lambda_{l  m}=||\delta^i_l \delta^j_m ||$.
The matrices have one label $l$ for the brane where the string leaves 
from (``start brane'') 
and one label $m$ for the brane where the string ends (``end brane'').

When computing an open string amplitude with Chan-Paton factors 
we first make a product of Chan-Paton matrices and this
ensures that the end brane of a string is the start
brane of the next one, then we take the trace of this product
so that the brane where the last string ends is the brane where the
first string originates. The result of the trace accounts for the degeneracy
of diagram one is considering but in the generic setting this
degeneracy is simply one therefore using the naive Chan-Paton factors in
computing an amplitude involving $N$ branes in a generic position
would yield the wrong degeneracy. 
In our case where the branes are in a generic position the role of the
two labels can be taken by $y_0$ for the start brane and
$y_\pi=y_0+2 \pi (w+\theta) R'$ for the end brane.
In computing amplitudes we have now to take care that the end brane of
a string is the start brane of the next one since this is not anymore
implied by matrix multiplication.
The effect of momentum (actually width) conservation in DD
directions in this formalism is to ensure that the sum of all %signed
string widths, defined each as $y_\pi-y_0$, is zero: this is actually
not a new constraint but a consistency condition.
% since this is
%implied by the fact the the end brane of a string is the start brane
%of the next one.

In the limit where some branes are superimposed, the brane
position is not anymore enough to distinguish among them and it must
be supplemented with a label, thus recovering Chan-Paton factors.
\COMMENTO{
In this expression Chan-Paton factors are not used since we suppose
D-branes be in a generic, non overlapping, position (and hence we are
restrict to the case of Dp with $p\le25$) where the gauge
group is a product of $U(1)$ factors. 
Nevertheless it is possible both to find a remnant of Chan-Paton factors and
to recover them. 
In fact every open string %in presence of DD directions 
is characterized by a kind of Chan-Paton factor: $y_0$ and
$y_\pi=y_0+(w+\theta) R'$.
While a true ($U(N)$) Chan-Paton factor is nothing but a couple of labels: 
a label for the start brane plus a label for the end brane, in
our case where branes are in generic position these labels
can be traded for  ``the starting position'' $y_0$ and ``the
end position'' $y_\pi=y_0+(w+\theta) R'$.
Because of this in eq. (\ref{Ademampl}) one has to take care that open 
strings are
in the proper order so that the brane where one string ends is the
brane where the following string starts: the
``momentum'' conservation in DD directions takes care that the signed total
brane distance be zero.
The non generic situation with enhanced gauge symmetry is recovered
when more branes are superimposed in which case for any superimposed
string the information on start and end brane becomes a true
Chan-Paton factor.
} % END COMMENTO

It is interesting to note that amplitude (\ref{Ademampl}) is 
well defined even when there are some non compact directions with DD
boundary conditions.
%only ill
%defined if $N_o=0$  in the uncompactified limit when all the $X$ have
%Neumann boundary condition (\cite{Ademetal}) while it is well 
%defined when the space-time is compact: it is in fact  the integration over
%the positions (when $N_c>1$) which is divergent but this can be avoided in
%compact space,
%in fact, in eq. \eq{Ademampl} 
%momentum conservation \footnote{For $N_o=0$  and all $X_{NN}$ the
%vertices zero modes implies a factor 
%$\langle 0|~\exp\Big[{\rm i}\,
%q \sum\limits_{j=1}^{N_c}\left(k_{j L}+k_{j R}\right)\Big]
%| 0\rangle=~\delta_{\sum\limits_j n_j\,,\,0}$
%for any compactified direction.} constrains only the Kaluza-Klein part
%of the momentum and leaves its winding number arbitrary thus
%the singularity is avoided when factorizing in the closed channel
%since winding can flow in the sewing propagator. 
%This has to be contrasted with the non compact case treated in .
%On the other hand when some $X$ have Dirichlet boundary condition 
%then the scalar product of both compact and non compact directions 
%is proportional to  $\langle0|0\rangle=1$ 
%On the other hand 
When some non compact $X$ have Dirichlet boundary condition
it would seem that the correlator in
(\ref{Ademampl}) be divergent since it is proportional to
$_{non~compact~DD}\langle0|0\rangle_{non~compact~DD}$
where $|0\rangle_{non~compact~DD}$ are
the would-be momentum vacua in the non compact Dirichlet directions.
These vacua  $|0\rangle_{non~compact~DD}$ should naively be normalized as 
$_{non~compact~DD}\langle0|0\rangle_{non~compact~DD} \propto \delta(0)$
since the corresponding directions are non compact
but this is not the case.
In fact we can normalize
$_{non~compact~DD}\langle0|0\rangle_{non~compact~DD}=1$ as 
$|0\rangle_{non~compact~DD}$ is the 
only finite energy state because all the ``winding states'' have
energy proportional to $R'\rightarrow\infty$.
This  justifies the point of view that non compact DD directions must
be understood as a limit of compactified ones, in particular
the spectral decomposition of unity is then given by
\begin{equation}
\uno_{\mbox{non compact DD}}=\lim_{R'\rightarrow\infty}\sum_w
|\frac{w R'}{\alpha'} \rangle\langle \frac{w R'}{\alpha'}|
\end{equation}
where $|\frac{w R'}{\alpha'} \rangle$ are the eigenstates of the
corresponding $p_{non~compact~DD}$.

\section{The cocycles for vertices.}
\label{par2}

In this section
we first determine the cocycles for closed string vertex operators in closed
string formalism so that two closed string vertices commute with each other,
then we fix the cocycles for closed string vertex operators 
in open string formalism by demanding  that the normal ordering of the product
%OPE 
of two
closed string vertices be the same in both closed and open string
formalism: this is done 
 in order to be sure that it is possible to describe closed
string  emission in open formalism at all mass levels and that
open/closed string amplitudes are correctly and consistently factorizable 
in the closed channel.
We fix also the cocycles for open string emission vertices so that
they commute with closed string ones. In doing so we realize that also
$y_0$, the quantity entering the left and right moving part of the
open string fields given in eq.s (\ref{XL-XR}), 
has to change after the emission of an open string from the
$\sigma=0$ border and this change is fundamental even when open
string has NN boundary condition (see section \ref{explicit-example}
for an explicit example). 

All these cocycles and the transformation of $y_0$ turn out to be
conceptually very important for a 
proper amplitudes factorization but of little practical importance as
long as one is willing to drop phases in amplitudes.
%%%%%%%%%%%%%%%%%%%%%%%%%%%%%%%%%%%%%%%%%%%%%%%%%%%%%%%%%%%%%%%%%%%%%%
%%%%%%%%%%%%%%%%%%%%%%%%%%%%%%%%%%%%%%%%%%%%%%%%%%%%%%%%%%%%%%%%%%%%%%

\subsection{The cocycles for the closed string.}
\label{clos_stri_coc_sub}
As said before the first step is to determine the cocycles 
in the closed string formalism. 
A well defined CFT for closed strings must satisfy at least the
following criteria
\begin{enumerate}
\item closed string vertices commute;
\item a proper behavior under Hermitian conjugation.
\end{enumerate}
The fist request is necessary in order to ensure the mutual locality of closed
vertices (\cite{PolBook}), or in other words, that vertices obey the 
 spin-statistics theorem.
Usually vertices are written without cocycles as
\begin{eqnarray}
{W}^{(c)}_{\beta_L,\beta_R}(z,\bar z; k)
&=&
V_{\beta_L}(z;k_L)
\,\, {\tilde V}_{\beta_R}({\bar z};k_R)
\end{eqnarray}
where the right moving ${\tilde V}_{\beta_R}$ is a normal ordered
functional of the closed string right moving part
\begin{equation}
X^{(c) \mu }_{R}(\bar z) 
  =  
\tilde q^{\mu}_R
  -{\rm i}\, {2\alpha'}\, \tilde p^{ \mu}_R{}\ln {\bar z} +
  {\rm i}\,\sqrt{2\alpha'}\,
  \sum_{n\not =0} {sgn(n)\over {\sqrt {|n|}}} \tilde a_{R n}^{\mu}{\bar z}^{-n}
\end{equation}
and similarly for the left moving part
\begin{equation}
X^{(c) \mu }_{L}( z) 
  =  
 q^{\mu}_R
  -{\rm i}\, {2\alpha'}\,  p^{ \mu}_L{}\ln { z} +
  {\rm i}\,\sqrt{2\alpha'}\,
  \sum_{n\not =0} {sgn(n)\over {\sqrt {|n|}}}  a_{L n}^{\mu}{ z}^{-n}
\end{equation}
with $z\in \complessi$ and 
\begin{equation}
[q_L^\mu, p_L^\nu]=[q_R^\mu, p_R^\nu]= i \eta^{\mu \nu}
~~~~
[a_m^\mu, a^{\dagger \nu}_n]= 
[\tilde a_m^\mu, \tilde a^{\dagger \nu}_n]=\delta_{m,n} \eta^{\mu \nu}~~
m,n>0
\end{equation}
The previous expansions are however not completely exact for the non
compact directions since 
in  this case  the zero modes are common to
left and right moving sectors. 
Because of this the non compact left and right part of the vertices  are not
separately normal ordered but
only the non zero modes parts are normal ordered separately while
the common zero modes are normal ordered together.

For implementing the wanted properties we look for a solution of the
form (\cite{PolBook})
\begin{eqnarray}
\label{W_closed}
{\cal W}^{(c)}_{\beta_L,\beta_R}(z,\bar z; k)
&=&
c(k_L,k_R;p_L,p_R) V_{\beta_L}(z;k_L)
\,\, {\tilde V}_{\beta_R}({\bar z};k_R)
\end{eqnarray}
where the cocycles are given by
\begin{eqnarray}
\label{c_closed}
c(k_L,k_R;p_L,p_R)&=&
c_L(k_L,k_R; p_L)\, c_R(k_L,k_R; p_R)
%e^{i\pi\alpha'\left[(b k_L + c k_R)_a\, p_L^a
%+(d k_L + e k_R)_a\, p_R^a \right]}
\nonumber\\
&=&
e^{i\pi\alpha'(\bb k_L + \cc k_R)_a\, p_L^a} \,
e^{i\pi\alpha'(\dd k_L + \ee k_R)_a\, p_R^a}
~~
\label{cocycl-closed}
\end{eqnarray}
%$z\in \complessi$.
We choose the  coefficients $\bb,\cc,\dd,\ee$ to be matrices in order to be
general. 
They  have possibly only non vanishing entries in compact directions, 
e.g $\bb_a^{.b}$,  since only when there are compact directions they are 
 needed.
They must be determined, as said
before, so that two arbitrary vertices are mutually local, i.e. commute. 
This is also equivalent to the fact that the radial
ordering of a product of vertices is given by a unique expression
which can be derived by %appropriately 
analytically continuing whichever
particular ordering of the vertices is chosen to
perform the computation.

In order to compute these matrices we consider the ordering of
the product 
%OPE 
of two arbitrary vertices as
\begin{eqnarray}
\label{OPE_closed}
{\cal W}^{(c)}_{\beta_L,\beta_R}(z,\bar z; k_1)
{\cal W}^{(c)}_{\alpha_L,\alpha_R}(w,\bar w; k_2)
&=&
e^{i\Phi_{(c)}(k_1,k_2)}
c(k_{L1}+k_{L2},k_{R1}+k_{R2}; p_L,p_R) 
\nonumber\\
&&\times
V_{\beta_L}(z;k_{L1})
V_{\alpha_L}(w;k_{L2})
\,\, \, \,
%\nonumber\\
%&&
{\tilde V}_{\beta_R}({\bar z};k_{R1})
{\tilde V}_{\alpha_R}({\bar w};k_{R2})
\nonumber\\
\end{eqnarray}
where we have defined the phase
\begin{equation}
\label{OPE_phase_closed}
\Phi_{(c)}(k_1,k_2)
=
-\pi\, \alpha'\left[
\left( \bb\, k_{L2} +\cc \, k_{R2}\right) \cdot k_{L1} 
+\left( \dd \, k_{L2} +\ee \, k_{R2}\right) \cdot k_{R1} 
\right]
\end{equation}
and then we use  the well known formula 
($-\pi\le arg(z),arg(w)\le+\pi$, $|z|<|w|$) (see also appendix \ref{app_SDS})
\begin{equation}
\left[V(z;k_1)\,V(w;k_2)\right]_{\mbox{an.con.}}
= V(w;k_2)\,V(z;k_1)
e^{+i\pi\,sgn(arg(z)-arg(w))\,2\alpha'\, k_1 \cdot k_2}
\end{equation}
to compare with the product of the vertices in the opposite order
 $ 
{\cal W}^{(c)}_{\alpha_L,\alpha_R}(w,\bar w; k_2)
{\cal W}^{(c)}_{\beta_L,\beta_R}(z,\bar z; k_1)
$
and determine the unknown matrices.
From the previous equations, as it is described in 
appendix (\ref{details_Cstr}), 
we can deduce that the (matrix) coefficients in compact directions are
\begin{equation}
\label{CS_coeff}
\cc=-\dd=diag(1+2N_a)
\,\,\,\,
\bb=-\ee= 2~diag(M_a)
\end{equation}
where $N_a,M_a\in\interi$.
In the rest of the paper we choose
\begin{equation}
\cc=-\dd=\uno,
~~~~\bb=-\ee= 0
\end{equation}
so that the cocycle in eq. (\ref{c_closed}) simply becomes\footnote{
It is also possible to choose
$c(k_L,k_R;p_L,p_R)
=
e^{i \pi\alpha'( k_L + k_R)_a (p_L-p_R)^a}
=
e^{i \pi n_a \hat w^a}
$
where $\hat w^a$ is the winding operator if one is willing to have
vertices which are hermitian up to a phase and, hence, have a
Zamolodchikov metric not equal to the unity.}
\begin{equation}
c(k_L,k_R;p_L,p_R)
=
e^{i \pi\alpha'( k_{R a} p_L^a - k_{L a} p_R^a) }
\end{equation}

The action of T-duality on the cocycle of the vertex is not trivial,
since it exchanges the possible solutions available, 
i.e.  a T-duality along the compact direction $a$ gives  
$(\cc_{a a},\dd_{a a}) \leftrightarrow (\dd_{a a},\cc_{a a})$,
$(\bb_{a a},\ee_{a a}) \leftrightarrow (\ee_{a a},\bb_{a a})$ or
$N_a \leftrightarrow -N_a-1$, $M_a \leftrightarrow -M_a$.
The question is now whether we have different theories since different
choices of cocycles yield amplitudes which differ by a phase.
This question assumes that we read the actual effective theory vertices, signs
included, from the string amplitudes.
At first look we would answer yes since there is no way of redefining
the fields of the effective action 
in a such a way that phases in front of  amplitudes match,
as it is easy to verify in the case of the three tachyons (see
eq. (\ref{3T}) in appendix).
On the other side also loop corrections are proportional to the same
phase of the tree level since moving the cocycles in front of the
string of vertices yields the same numeric phase as in the tree level
case (this phase depends on external
momenta only and does not depend on the number of handles)
times an operatorial cocycle like 
$e^{i \pi \alpha' ~(\bb \sum_{j=1}^{N_c} k_{L j}+
  \cc \sum_{j=1}^{N_c} k_{R j}) \cdot p_L + \dots }$.
Now the above operatorial cocycle is identically $1$ since
the numeric coefficients are zero because they are 
proportional to the sum of the
external momenta which are conserved. 
Because of this the would-be different theories 
we get give different  S-matrix elements 
which nevertheless yield the same transition probability to all orders in
perturbation theory.
Moreover the physical states created by the different vertices are
the same.

%%%%%%%%%%%%%%%%%%%%%%%%%%%%%%%%%%%%%%%%%%%%%%%%%%%%%%%%%%%%%%%%%%%%%%
%%%%%%%%%%%%%%%%%%%%%%%%%%%%%%%%%%%%%%%%%%%%%%%%%%%%%%%%%%%%%%%%%%%%%%

\subsection{The cocycles and $y_0$ shift in the open string formalism.}
A well defined CFT for an open-closed string theory must satisfy
at least the following 5 constraints:
\begin{enumerate}
\item
the open string vertices at $\sigma=0$ and $\sigma=\pi$ commute;
\item
the open string emission vertex from $\sigma=0$ commutes with the
closed string vertices;
\item
in a similar way the open string emission vertex from $\sigma=\pi$
commutes with the closed string vertices;
\item
the closed string vertices in open string formalism must have the same
``OPE'' as in closed string formalism, or more precisely we want the
normal ordering of the product of two closed string vertices be
formally equal to (\ref{OPE_closed}), in particular
we want the closed string vertices to commute;
\item a proper behavior under Hermitian conjugation.
\end{enumerate}

The previous constraints can fail only because of phases and these
phases depend only on momenta therefore the problem amounts to find
the proper cocycles and, as we discuss later, the proper shift of the
$y_0$ (which enters the definition of left and right moving part of the
open string expansion given in eq.s (\ref{XL-XR})).

Before writing the vertices 
it is worth stressing once again what discussed in section \ref{par1}: 
we want the closed string vertices be expressed using the open string
operators only, i.e. closed strings are made of the same ``stuff'' of
the open string they are emitted.
We derive the vertices in appendix \ref{details_Ostr}, here we limit ourselves
to write them explicitly
\begin{eqnarray}
{\cal W}_{\beta_L,\beta_R}(z,\bar z; k_L,k_R)
&=&
%e^{-i\frac{1}{2}\pi\alpha'\,(k_L^2 - k_R^2) }
e^{-i\pi\alpha'\, (S k_L \cdot k_R )}
{ V}_{\beta_L}(X_L(z);k_L)
{}~{V}_{\beta_R}(X_R(\bar z); k_R)
\nonumber\\
\label{VV-0}
\\
{\cal V}_{\alpha}(x; k)
&=&
{ V}_{\alpha}(X_L(x)\YZp;k)
\,~~~~x>0
\label{V-0}
\end{eqnarray}
along with the vertices for the emission from the $\sigma=\pi$ boundary
\begin{eqnarray}
{\cal V}_{\alpha}(x; k)
&=&
e^{-2i\pi\, k\cdot p}
{ V}_{\alpha}(X_L(x)\YZp;k)
\,~~~~x<0.
\label{V-pi}
\end{eqnarray}
In this case the action of the cocycle simply means that we can substitute 
$\left[X_L(x)-y_0\right]_{zero~mode}=q-2\alpha' i p \log(x) \rightarrow
q-2\alpha' i p \log( |x |)$ and drop the cocycle.

While the previous vertices seem essentially the normal vertices, the
closed string vertex does depend on $y_0$ since
%must be supplemented with a change of $y_0$ according to the open
%string from which the previous states are emitted from. 
%The reason why we pay attention to $y_0$ is that 
\begin{equation}
{\cal W}_{\beta_L,\beta_R}(z,\bar z; k_L,k_R)
\sim e^{i y_0 \cdot (k_L -S k_R)} \dots
\end{equation}
and $y_0$ changes along the worldsheet.
This happens
because we are taking the original point of view of \cite{Ademetal} 
according to which the ``stuff'' of the ``carrier'' string may change after 
the emission of another open string. 
Because of this the closed string emission vertices do not change in
form when written using the ``carrier'' string fields from which the
closed string states are emitted 
but they do change when the actual ``carrier'' string
fields are expressed  using the fields of incoming ``carrier'' string.

\begin{figure}[hbtp]
\begin{center}
\begin{minipage}[t]{0.45\linewidth}
\scalebox{0.9}{
\includegraphics[type=eps,ext=.eps,read=.eps,width=0.9\textwidth]{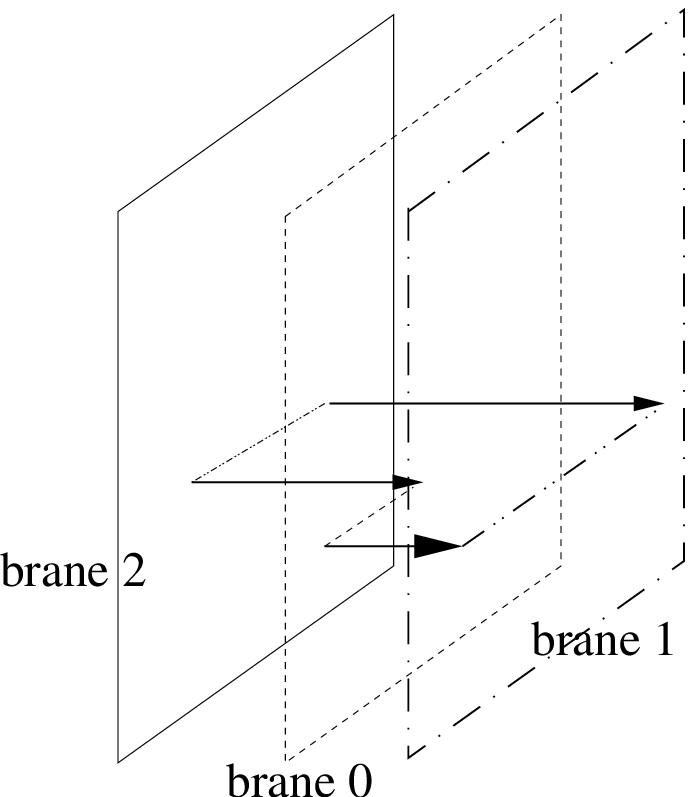}
}
\end{minipage}
\begin{minipage}[t]{0.45\linewidth}
\scalebox{0.9}{
\includegraphics[type=eps,ext=.eps,read=.eps,width=0.9\textwidth]{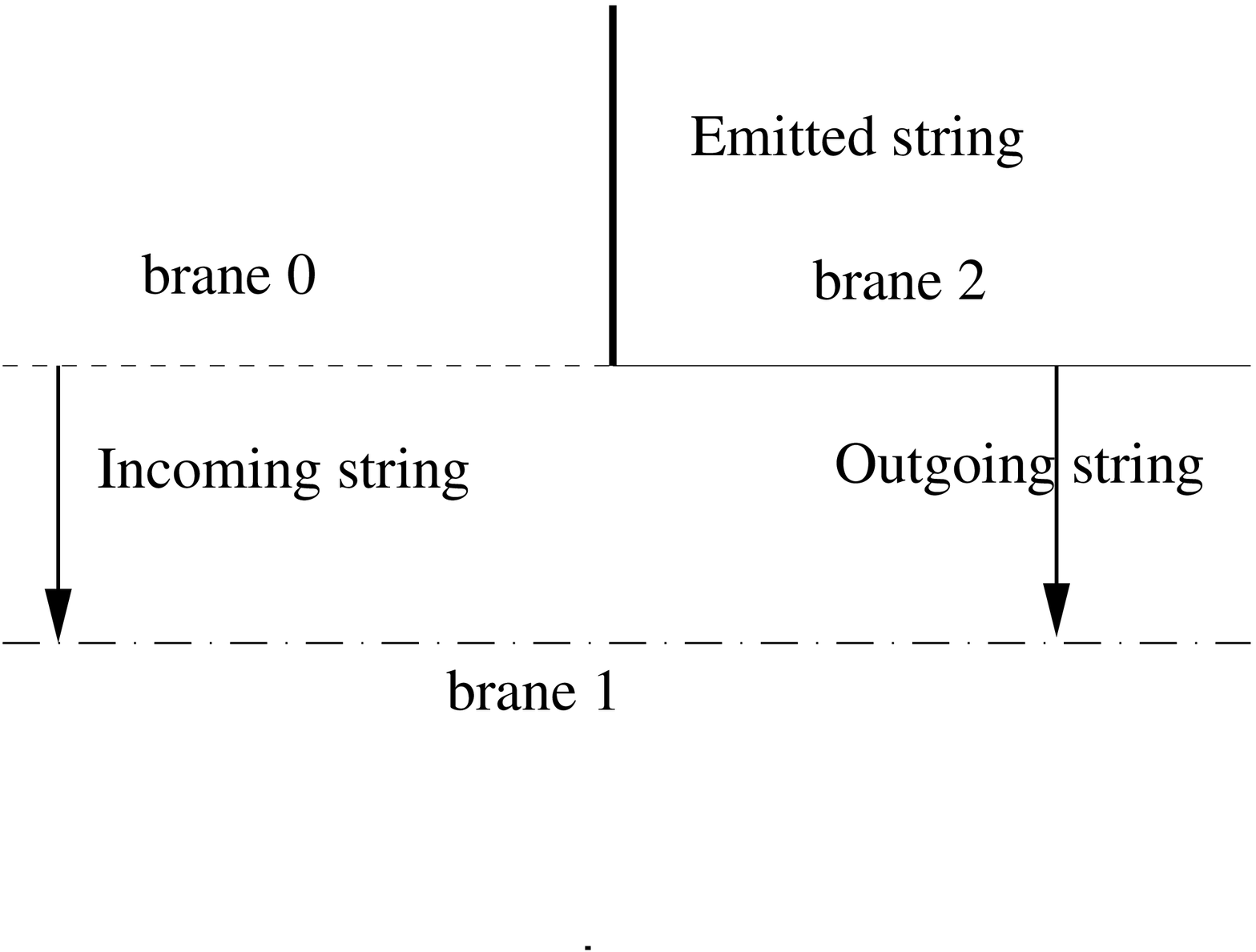}
}
\end{minipage}

 \caption{
Two different pictures of the emission of a string from an incoming
one which transforms into an outgoing string.
The incoming string stretches between brane 0 at $y_0^{in}$ 
and brane 1 at $y_0^{in}+\delta_{in}$ ($\delta_{in}=2\pi\alpha' k_{in}$);
the emitted string stretches between brane 2 at 
$ y_0^{out}$ and brane 0 at
  $y_0^{out}+ \delta_{emitted} = y_0^{in}$
($\delta_{emitted}=2\pi\alpha' k_{emitted}$) 
and the outgoing string between brane 2 
at $y_0^{out}= y_0^{in}- \delta_{emitted} $ and brane 1 at
$y_0^{out}+\delta_{emitted}+\delta_{in}$.
\label{fig:y0-recoil}
}
\end{center}
\end{figure}

The change  of $y_0$ can be
better appreciated by considering a carrier open string with Dirichlet
b.c. in at least one compact direction.
In fig.s \ref{fig:y0-recoil} we have pictured in two ways the
interaction between an incoming open string stretched between brane 0
and brane 1 with another open string, the emitted string, stretched
between brane 2 and brane 0.
The incoming string has DD b.c. $X^{in}(\sigma=0)=y_0$ and 
 $X^{in}(\sigma=\pi)=y_0+ \delta^{in}$ with 
$\delta^{in}=2\pi \alpha' k_{in}$.
The other string, which is emitted from the incoming one, stretches 
between  
$X^{emitted}(\sigma=0)=y_0^{emitted}=y_0-\delta^{emitted}$
and
$X^{emitted}(\sigma=\pi)=y_0^{emitted}+\delta^{emitted}=y_0$.
The result of the interaction is the outgoing string
with
$X^{out}(\sigma=0)=y_0^{out}=y_0^{emitted}=y_0-\delta^{emitted}$
and
$X^{out}(\sigma=\pi)=y_0^{out}+\delta^{out}=y_0+\delta^{in}+\delta^{emitted}$.
The latter equation has a simple interpretation:
when a string with DD b.c. is emitted from the $\sigma=0$ border of
the incoming open string 
the outgoing string has width equal to the sum of the incoming carrier string
$\delta_{in}$  and the emitted string $\delta_{emitted}$.
This is taken care by the momentum $p$ conservation but
the position of the $\sigma=0$ border of the outgoing string 
is {\sl changed} in such a way the 
position of the $\sigma=\pi$ border is left unchanged, i.e. 
\begin{equation}
y_0^{out}=y_0-\delta_{emitted}
~~~
\delta_{emitted}=2\pi \alpha' k_{emitted}
.
\end{equation}

The main point hence is that the open string changes its b.c. 
after the emission of a string from the $\sigma=0$ boundary
and therefore $y_0$ changes too accordingly.
Hence the closed string vertex does change  when
expressed using the incoming string fields.
%in the sequence of vertices entering the desired correlator.

Because of the expansions in eq.s (\ref{XL-XR}) this change in $y_0$ means 
\begin{equation} 
X^{out}_L=X^{in}_L-\delta_{emitted}
\end{equation}
and
\begin{equation} 
X^{out}_R=X^{in}_R+ S \delta_{emitted}
\end{equation}
Hence the closed string vertex changes when
inserted before or after the emission of the open string from the
$\sigma=0$ boundary.
The proper
commutation relation between a closed string and an open string
emitted from $\sigma=0$ then reads
\begin{eqnarray}
\hspace{-5em}
\left[
{\cal W}_{\beta_L,\beta_R}(X_L^{out}(z),X_R^{out}(\bar z))
\, 
{\cal V}_{\alpha}(x; k)
\right]_{an. con.}
&=&
{\cal V}_{\alpha}(x; k)
\,
{\cal W}_{\beta_L,\beta_R}(X_L^{in}(z),X_R^{in}(\bar z))
\nonumber\\
&=&
{\cal V}_{\alpha}(x; k)
\,
{\cal W}_{\beta_L,\beta_R}(X_L^{out}(z),X_R^{out}(\bar z))
e^{2\pi \alpha' i\,k\cdot(k_L-Sk_R)}
\nonumber\\
&=&
\left[
{\cal W}_{\beta_L,\beta_R}(X_L^{in}(z),X_R^{in}(\bar z))
\, 
{\cal V}_{\alpha}(x; k)
\right]_{an. con.}
e^{-2\pi \alpha' i\,k\cdot(k_L-Sk_R)}
\nonumber\\
\label{W_V_commutation}
\end{eqnarray}
where we have used the fact that $\delta=2\pi \alpha' k$ and $an.con.$
means analytically continued.

\subsection{An explicit example.}
\label{explicit-example}
The importance of the cocycle factors and the $y_0$ shift  for a full
 consistency of amplitudes  
and the proper understanding of the consequences of commutation
relation eq. (\ref{W_V_commutation})
can be appreciated computing in the operatorial formalism
the one closed tachyon - one open tachyon amplitude in two different
ways either
\begin{equation}
\langle 0| {\cal W}_T(k_L,k_R;z,\bar z)\, {\cal V}_T(k;x)|0\rangle
~~~~~~ |z| > |x|
\end{equation}
or
\begin{equation}
\langle 0|{\cal V}_T(k;x) \, {\cal W}_T(k_L,k_R;z,\bar z) |0\rangle
~~~~~~ |x| > |z| .
\end{equation}
In the second way we find
\begin{eqnarray}
&&
%e^{i\frac{1}{2}\pi\alpha'\,(k_L^2 - k_R^2) }
~\langle 0|{\cal V}_T(k;x) \, {\cal W}_T(k_L,k_R;z,\bar z) |0\rangle
~~~~~~ |x| > |z| 
\nonumber\\
&&
\langle0| :e^{i k \cdot (X_L(x)-y_0)}: ~ 
:e^{i k_L \cdot X_L(z)}: \, :e^{i k_R \cdot X_R(\bar z)}: 
 |0\rangle
\nonumber\\
&=&
e^{i (k_L-S k_R)\cdot y_0}
|z-\bar z|^{2\alpha' k_L\cdot Sk_R}
(x-z)^{2\alpha' k\cdot k_L}
(x-\bar z)^{2\alpha' k\cdot k_R}
\nonumber\\
&&(2\pi)^{D_{NN,nc}} \delta^{D_{NN,nc}}(k+k_L+k_R)
\delta^{D_{NN,c}+D_{DD}}_{k+k_L+S k_R,0}
\nonumber\\
&=&
e^{i (k_L-S k_R)\cdot y_0}
|z-\bar z|^{2\alpha' k_L\cdot Sk_R}
|x-z|^{2\alpha' k\cdot(k_L+Sk_R)}
\nonumber\\
&&\left(\frac{x-z}{x-\bar z}\right)^{\alpha' k\cdot(k_L-Sk_R)}
(2\pi)^{D_{NN,nc}} \delta^{D_{NN,nc}}(k+k_L+k_R)
\delta^{D_{NN,c}+D_{DD}}_{k+k_L+S k_R,0}
\nonumber\\
&=&
e^{i (k_L-S k_R)\cdot y_0 }
|z-\bar z|^{-1} |x-z|^{-2}
\left( 2\pi \delta(k+k_L+k_R) \right)^{D_{NN,nc}}
\delta^{D_{NN,c}+D_{DD}}_{k+k_L+S k_R,0}
\label{1Tc-1To}
\end{eqnarray}
where $D_{DD}$  is the number of directions  with DD boundary condition,
$D_{NN,c}$ ($D_{NN,nc}$) is the number of compact (non compact)
directions  with NN boundary condition and we have written the modulus
$|z-\bar z|^{\cdots}$ and not $(z-\bar z)^{...}$ 
because of the contribution from the cocycle $e^{-i\pi\alpha'\, (S k_L
  \cdot k_R )}$ of 
closed string vertex cocycle.
The last line is obtained using the momentum conservation and mass
shell conditions $k_L^2=k_R^2=k^2=\frac{1}{\alpha'}$ and shows how the
amplitude is conformal invariant when ghost contribution is added.

Let us now consider the computation for $|z|>x$. According to our
previous discussion it requires the use of the closed string emission
vertex depending on $X_{out}$%
\footnote{
Notice that even for DD directions where $X_{zero\,
  modes}=y_0-{\rm i}\, \sqrt{2\alpha'}\, a_0^{\mu}{}\ln \sqrt{z\over
  \bar z}$ 
for closed string vertices
there is a non trivial $z$ dependent normalization factor coming from
the splitting into left and right moving part of the string
coordinates, explicitly:
\begin{eqnarray*}
e^{ik_L(x+y_0-2\alpha' i p\ln(z))}
e^{-ik_R(x-y_0-2\alpha' i p\ln(\bar z))}
&=&
e^{i(k_L+k_R)y_0}
e^{-2\alpha' k_L\, k_R \ln(z)}
e^{2\alpha' p( k_L\ln(z)-k_R\ln(\bar z))}
e^{i(k_L-k_R)x}
\end{eqnarray*}
}
\begin{eqnarray*}
{\cal W}_T(k_L,k_R;z,\bar z)
&=&
e^{-i\pi\alpha' S k_L \cdot k_R}~
:e^{i k_L\cdot X_L^{out}(z)}:
\,
: e^{i k_R\cdot X_R^{out}(\bar z)}:
\\
&=&
e^{-i\pi\alpha' S k_L \cdot k_R }
~:e^{i k_L\cdot (X_L(z)-2\pi\alpha' k)}:
\,
:e^{i k_R\cdot (X_R(\bar z)+2\pi\alpha' S k )}:
\end{eqnarray*}
%where we have only  written $e^{-i\pi\alpha'[\dots]}$ since we want
%to stress the role of 
The shift of $y_0$  by $2\pi\alpha' k$ is necessary in computing
this correlator, in fact
\begin{eqnarray}
&&
%e^{i\frac{1}{2}\pi\alpha'\,(k_L^2 - k_R^2) }
~
\langle0|{\cal W}_T(k_L,k_R;z,\bar z)\, {\cal V}_T(k;x)0|\rangle
~~~~~~ |z| > |x|
\nonumber\\
&=&
\langle0| 
:e^{i k_L \cdot X_L^{out}(z)}: \, :e^{i k_R \cdot X_R^{out}(\bar z)}: 
:e^{i k \cdot (X_L(x)-y_0)}: ~ 
 |0\rangle
\nonumber\\
&=&
e^{-i 2\pi \alpha' k\cdot( k_L -S k_R) }
e^{i (k_L-S k_R)\cdot y_0}
|z-\bar z|^{2\alpha' k_L\cdot Sk_R}
(z-x)^{2\alpha' k\cdot k_L}
(\bar z-x)^{2\alpha' k\cdot k_R}
\nonumber\\
&&(2\pi)^{D_{NN,nc}} \delta^{D_{NN,nc}}(k+k_L+k_R)
\delta^{D_{NN,c}+D_{DD}}_{k+k_L+S k_R,0}
\nonumber\\
&=&
e^{i (k_L-S k_R)\cdot y_0}
|z-\bar z|^{2\alpha' k_L\cdot Sk_R}
|z-x|^{2\alpha' k\cdot(k_L+Sk_R)}
\nonumber\\
&&
e^{-i 2\pi \alpha' k\cdot( k_L -S k_R) }
\left(\frac{z-x}{\bar z-x}\right)^{\alpha' k\cdot(k_L-Sk_R)}
(2\pi)^{D_{NN,nc}} \delta^{D_{NN,nc}}(k+k_L+k_R)
\delta^{D_{NN,c}+D_{DD}}_{k+k_L+S k_R,0}
\nonumber\\
&=&
e^{i (k_L-S k_R)\cdot y_0}
|z-\bar z|^{2\alpha' k_L\cdot Sk_R}
|x-z|^{2\alpha' k\cdot(k_L+Sk_R)}
\nonumber\\
&&
\left(\frac{x-z}{x-\bar z}\right)^{\alpha' k\cdot(k_L-Sk_R)}
(2\pi)^{D_{NN,nc}} \delta^{D_{NN,nc}}(k+k_L+k_R)
\delta^{D_{NN,c}+D_{DD}}_{k+k_L+S k_R,0}
\nonumber\\
\end{eqnarray}
where the phase from the shift $e^{-i 2\pi \alpha' k\cdot( k_L -S k_R) }$
is necessary
to compensate the phase obtained analytically continuing  (also off
shell where $ k\cdot(k_L-Sk_R) \ne -(k_L+S k_R)\cdot(k_L-Sk_R) $ )
$\left(\frac{z-x}{\bar z-x}\right)^{\alpha' k\cdot(k_L-Sk_R)}$ to the
region $x>|z|$. In performing this analytical continuation one has
to use 
\begin{equation}
\left[ \log (z-w )\right]_{an. ~cont.} 
= 
\log (w-z) + i\pi~ sign(arg(z)-arg(w))
~~~~
-\pi< arg(z),arg(w)<\pi
\end{equation}
which in this case gives 
$\left[ \log (z-x)\right]_{an. ~cont.} = \log (x-z) + i\pi$ 
and its complex conjugate expression
$ \left[ \log (\bar z-x)\right]_{an. ~cont.} =\log(x-\bar z)-i \pi$.

%%%%%%%%%%%%%%%%%%%%%%%%%%%%%%%%%%%%%%%%%%%%%%%%%%%%%%%%%%%%%%%%%%%%%%%%%%%%%%
%%%%%%%%%%%%%%%%%%%%%%%%%%%%%%%%%%%%%%%%%%%%%%%%%%%%%%%%%%%%%%%%%%%%%%%%%%%%%%
%%%%%%%%%%%%%%%%%%%%%%%%%%%%%%%%%%%%%%%%%%%%%%%%%%%%%%%%%%%%%%%%%%%%%%%%%%%%%%
%%%%%%%%%%%%%%%%%%%%%%%%%%%%%%%%%%%%%%%%%%%%%%%%%%%%%%%%%%%%%%%%%%%%%%%%%%%%%%

\section{Amplitudes factorization. }
\label{par3}
Since we are trying to ensure that all intermediate expressions be
well defined with respect to vertices commutativity property and 
to keep track of how and where phases arise, 
the naive procedure
adopted in (\cite{Ademetal},\cite{Frau:1997mq}) is not completely
valid: details on how this can be accomplished are given in app. 
\ref{App:ChangeVar}, here we give only the main steps.

We start considering the correlator (which is denoted by a capital $A$
and is not the amplitude which
needs an integration over the moduli space which we consider later )%($x_{N_o+1}=+\infty$)
\begin{equation}
\label{corr-z}
A(N_o, N_c)= \prod_{i=1}^{N_o-1}\theta(|x_{i+1}|- |x_i|)\,
\langle 0 |~{\rm T}\left(
 \prod_{i=1}^{N_o} {\cal V}_{\alpha_i}(x_i;k_i)dx_i 
\,\prod_{j=1}^{N_c} {\cal W}_{\beta_{L j},\beta_{R j}}(z_j, \bar z_j;
k_{L/R ~j}) d^2z_j
\right)
| 0\rangle~~
\end{equation}
where the $x$s are real (both positive and negative),
all $z$s are in the upper half-plane $\,\,\Hplus$, 
the vacuum state $|0\rangle$ is defined as 
$|0\rangle=\prod_{\mu=1}^{\mu=D } |p_\mu=0;0_a\rangle$
%%%
and it is normalized as
\begin{equation}
\label{vacua_norm}
\langle p_\mu=0|p_\mu=0\rangle=\left\{ \begin{array}{cc} 
\mbox{non compact NN} & 2\pi\delta(0) \\
\mbox{non compact DD} & 1 \\
\mbox{compact} & 1 
\end{array}
\right.
\end{equation}
The normalization of the non compact directions with Dirichlet
boundary condition may seem strange but one has to keep in mind that
in this case only $|0\rangle_{\mbox{non compact DD}}$ 
has finite energy since the ``momentum'' spectrum is $\frac{w R}{\alpha'}$;
nevertheless the spectral decomposition of unity is still given by
\begin{equation}
\uno_{\mbox{non compact DD}}=\lim_{R\rightarrow\infty}\sum_w
|\frac{w R}{\alpha'} \rangle\langle \frac{w R}{\alpha'} |
\end{equation}
and because of this the non compact DD case must be understood as a special
decompactification limit: using the naive result would lead to a wrong result.
%Because of this from now on we consider non compact directions with
%NN boundary conditions.

\begin{figure}[hbtp]
\begin{minipage}[t]{0.45\linewidth}
\scalebox{0.9}{
\includegraphics[type=eps,ext=.eps,read=.eps,width=0.9\textwidth]{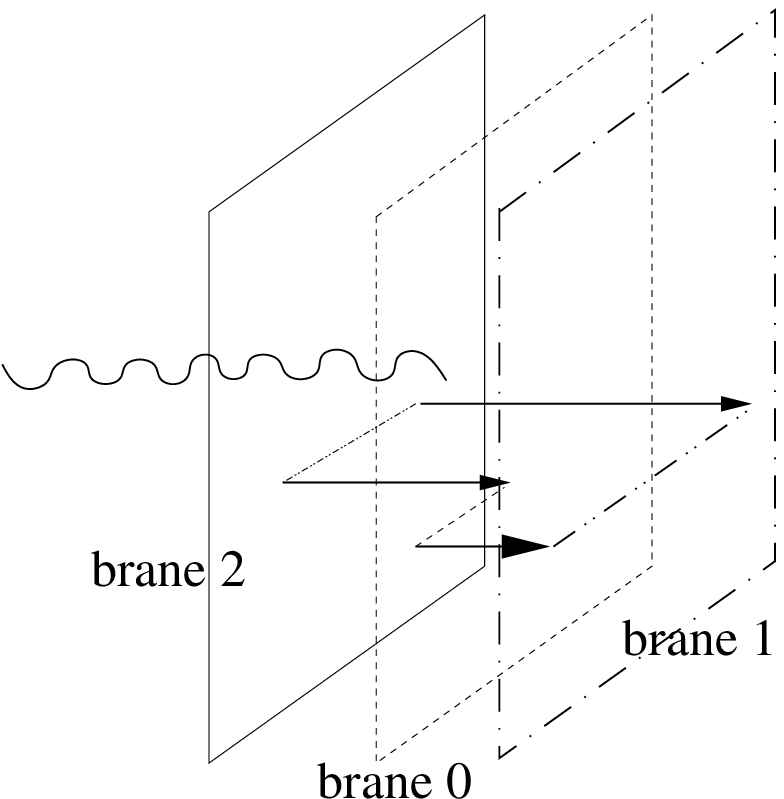}
}
 \caption{A 3 parallel branes interaction with 3 open strings and 1
 closed one.
 }
\label{figure:3open-1closed}
\end{minipage}
\hskip 1cm
\begin{minipage}[t]{0.45\linewidth}
\includegraphics[type=eps,ext=.eps,read=.eps,width=0.9\linewidth]{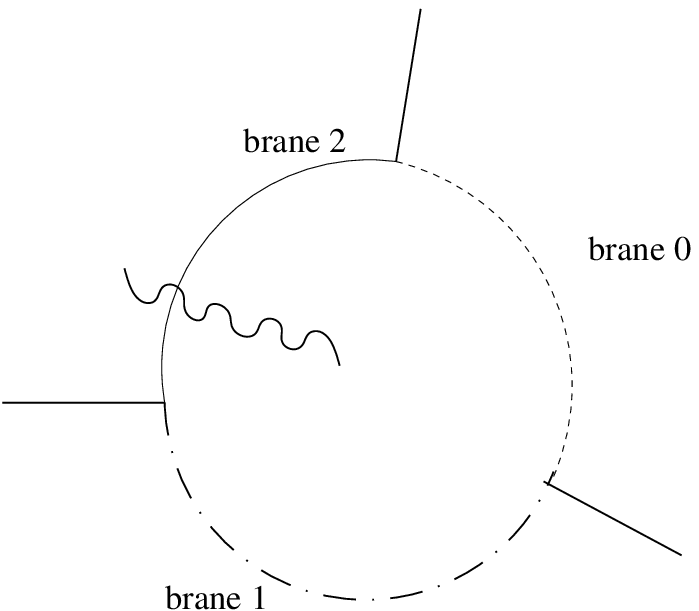}
 \caption{
The same configuration as the previous picture with the usual disk
picture. This representation is obtained after changing variables as 
in eq. (\ref{transf}).
}
\label{figure:disk-3open-1closed}
\end{minipage}

\end{figure}

On this correlator we 
want to perform the $SL(2,\complessi)$ transformation\footnote{
This is {\sl not} a symmetry of the amplitude.
}
\begin{equation}
\label{transf}
w={z+{\rm i} \over z-{\rm i}}~~~
\end{equation}
which maps $x\in\reali\rightarrow w(x)=e^{i\phi}$, $z\in \Hplus
\rightarrow w(z)$ with $|w(z)|>1$ and 
$\bar z \rightarrow w(\bar z)=(\overline{w(z)})^{-1}$
by inserting the corresponding $SL(2,\complessi)$ operator.
In this way we move form the graphical representation given in
fig. (\ref{figure:3open-1closed}) to the one in
fig. (\ref{figure:disk-3open-1closed}). 

%The naive procedure to implement this change of variable 
%has some small problems which are 
%discussed in app. (\ref{App:ChangeVar}) together with
%the right procedure. 
The final answer is  given by the following correlator as discussed  in
appendix \ref{App:ChangeVar} %($\phi_{N_o+1}=0$)
\begin{eqnarray}
\label{corr-wc0}
A(N_o, N_c)&=&
e^{i\alpha_0}
\langle 0 |~{\rm T}\Big(
 \prod_{i=1}^{N_o} \EYZi %e^{i p_{i} y_0} 
{ V}_{\alpha_i}(e^{i\phi_i};k_i)de^{i\phi_i}
\,\prod_{l=1}^{N_o-1}  \theta(\phi_{l}- \phi_{l+1})
\nonumber\\
&&
\,\prod_{j=1}^{N_o}
  c_L(k_{L j},k_{R j};p)\,e^{i k_{L j} y_0} 
  { V}_{\beta_{L j}}(w_j; k_{L j}) dw_j
\nonumber\\
&&
\,\prod_{l=1}^{N_c} 
  c_R(k_{L l},k_{R l};-Sp)\, e^{-i S k_{R j} y_0} 
 { V}_{\beta_{R l}}(\frac{1}{\bar w_l};
  k_{R l})
 \frac{d\bar w_l}{\bar w^2_l}
\Big)
| 0\rangle~~
\end{eqnarray}
where we have made explicit the $y_0$ dependence by redefining the
$X_{L,R}$ in such a way they are free of $y_0$ 
and we have introduced the cocycles $c_L$ and $c_R$ to ensure that
the amplitude can be obtained by  the analytic continuation of 
whichever radial order.
It is important to stress that thanks to the cocycle the phase $\alpha_0$
is independent of the different specific cases which arise from the 
radial ordering.

\subsection{Spectral decomposition of the unity in the disk channel.}

We can now proceed as in (\cite{ Ademetal},\cite{Frau:1997mq}) and  we insert a
complete set of states at radial time $1^-$. 
Since we are in the open string channel there is only one momentum
flowing for each direction and therefore we want to insert something like 
$\uno=\int_q |q\rangle\langle  q|$ as far as the zero modes are concerned.
The naive guess is that we
can insert the usual open string version of the spectral decomposition of unity
\begin{equation}
\uno_{naive}=\sum_{\lambda,q,n,w}|\lambda,(q,\frac{n}{R}
,\frac{w  R}{\alpha'} )\rangle \langle\lambda, (q,\frac{n}{R}
,\frac{w  R}{\alpha'} ) | 
\label{uno-naive}
\end{equation} 
with continuous momentum $q$ flowing in non compact NN directions,
 momentum $\frac{n}{R}$ flowing in compact NN directions 
and winding $\frac{w  R}{\alpha'}$ flowing in compact DD
ones, unfortunately this guess is wrong. 
To understand what is going on we analyze
what flows in compact directions. 
The momentum conservation reads
\begin{equation}
\sum_{j=1}^{N_c}(k_{L j} + S k_{R j}) + \sum_{i=1}^{N_o} k_{i}=0
\end{equation}
(with $S=+1$ in NN directions and $S=-1$ in DD directions)
and because we are inserting $\uno$ at  radial time $1^-$
this equation can be split into two different relations
\begin{equation}
\left\{\begin{array}{cc}
\sum_{j=1}^{N_c} S k_{R j} + q=0 
& \frac{1}{|w|}<1^-
\\
-q+\sum_{j=1}^{N_c} k_{L j} + \sum_{i=1}^{N_o} k_{i}=0
&  |e^{i\phi}|, |w|> 1^-
\end{array}\right.
\end{equation}
In compact NN  directions the second equation can be written in a
more explicit way as 
\begin{equation}
 \frac{1}{R}(\frac{1}{2}\sum_{j=1}^{N_c} n_j 
                   + \sum_{i=1}^{N_o}(n_{i}+\theta_i) )
+\frac{R }{2 \alpha'} \sum_{j=1}^{N_c} w_j 
-q=0
\label{443}
\end{equation}
(and in compact DD direction as
\begin{equation}
 \frac{1}{2 R} \sum_{j=1}^{N_c} n_j 
+\frac{R}{\alpha'}(
  \frac{1}{2} \sum_{j=1}^{N_c} w_j
+\sum_{i=1}^{N_o}(w_{i}+\theta_i) )
-q=0
\end{equation}
).
Since winding $w$ (momentum $n$) of the closed string states must be unrestricted in NN (DD) directions 
a first guess would be to sum over
$q=\frac{R  w}{2 \alpha'}$ $\forall w\in\Z$ 
($q=\frac{n}{2 R}$  $\forall n\in\Z$) but this is not yet
completely right.
Choosing these values for $q$ eq. (\ref{443}) for the NN directions becomes
\begin{equation}
\frac{1}{2}\sum_{j=1}^{N_c} n_j + \sum_{i=1}^{N_o}(n_{i}+\theta_i)=0
\end{equation}
and 
because of the factor $\frac{1}{2}$ which multiplies $\sum_{j=1}^{N_c} n_j$
in this equation, 
we get a wrong momentum conservation (and similarly for DD directions).
On the other hand summing over all $q=\frac{1}{2}\left(\frac{n}{R}+\frac{w
R}{\alpha'}\right)$  (both $n$ and $w$) would give a wrong one loop
amplitude.
The right answer for generic $R$s, even if strange, is the following 
 (closed string like) spectral decomposition of
unity\footnote{
Notice the use of the notation $\{\}$ and $()$ in the definitions of vectors.
}
\begin{eqnarray}
\label{spect_1}
\uno
&=&
\sum_{\lambda,q,w,n} |\lambda,\{q,w,n\}\rangle
\langle\lambda,\{q,w,n\}|
\nonumber\\
&=&
\sum_{\lambda,q,w,n}
|\lambda,( q
          ,-\frac{\sum_{i=1}^{N_o} k_{i}}{2}+\frac{w  R}{2\alpha'}
          ,-\frac{\sum_{i=1}^{N_o} k_{i}}{2}+\frac{n}{2 R} ) \rangle
\langle\lambda,( q
          ,-\frac{\sum_{i=1}^{N_o} k_{i}}{2}+\frac{w  R}{2\alpha'}
          ,-\frac{\sum_{i=1}^{N_o} k_{i}}{2}+\frac{n}{2 R} ) |
\nonumber\\
~~~
\end{eqnarray}
where 
we have a sum of shifted {\sl winding} for any NN compact direction 
and sum of shifted {\sl momentum} for any DD compact direction.
In non compact directions
$\sum_q$ has to be understood as the result of decompactification
limit, i.e.  $\int \frac{d q}{2\pi}$ for any
non compact DD direction and 
$\lim_{R\rightarrow\infty}\sum_w |\frac{w  R}{\alpha'}\rangle
\langle \frac{w  R}{\alpha'}|$
for any non compact NN direction.
This result is a clue that already at tree level open string knows of
the existence of closed string. %due to the exchange of $\sigma$ and
%$\tau$. and bears some resemblance with  (\cite{Sen:2003xs}).
On the other side one could argue that is not completely surprising
since we are computing a mixed open-closed amplitude, even so
we can understand better why this happens by noticing that
this spectral decomposition of unity is not what one naively would
expect because, while factorizing, we are separating the left and right part of
closed string vertex operators.
%and actually 
%the naive spectral decomposition (\ref{uno-naive}) would
%be the right one if we factorized an amplitude with only open strings
%emission or without splitting the left and right part of closed string
%vertices.

\subsection{Factorizing the disk amplitude.}

Inserting the unity given in eq (\ref{spect_1}) into
eq. (\ref{corr-wc0}) we get 
\begin{eqnarray}
\label{corr-wc1}
A(N_o, N_c)&=& 
e^{i\alpha_0}
e^{i y_0 \left( \sum_{j=1}^{N_c}(k_{L j} -S k_{R j}) 
               \YZS  \right)} 
\prod\theta(\phi_{i}- \phi_{i+1})
\nonumber\\
&&
\sum_{\lambda,q,w,n}
\langle 0 |
T\left(
\,\prod_{j=1}^{N_c} 
  c_L(k_{L j},k_{R j};p)\, 
{ V}_{\beta_{L j}}(w_j; k_{L j}) dw_j
\right)
\nonumber\\
&&
 \prod_{i=1}^{N_o} 
{\cal V}_{\alpha_i}(e^{i\phi_i};k_i)de^{i\phi_i}
|\lambda,\{q,w,n\}\rangle
\nonumber\\
&&
\langle\lambda,\{q,w,n\}|
T\left(
\,\prod_{l=1}^{N_c} 
  c_R(k_{L l},k_{R l};-Sp)\, 
{ V}_{\beta_{R l}}(\frac{1}{\bar w_l};
  k_{R l})
 \frac{d\bar w_l}{\bar w^2_l}
\right)
| 0\rangle~~
\nonumber\\
~~
\end{eqnarray}
We can then take the transpose of the piece containing $V_{\beta_R}$s
\footnote{
In performing this transposition  
we again have to pay attention to the cocycles which ensure that we get
the same phase in the transposed expression as in the original one.
}
and get
\begin{eqnarray}
\label{corr-wc2}
A(N_o, N_c)
&=& 
e^{i\alpha_1}
e^{i y_0 \left( \sum_{j=1}^{N_c}(k_{L j} -S k_{R j}) 
                \YZS \right)} 
\prod\theta(\phi_{i}- \phi_{i+1})
\nonumber\\
&&\sum_{\lambda,q,w,n}
\langle 0 |
T\Big(
\,\prod_{j=1}^{N_c} 
  c_L(k_{L j},k_{R j};p)\, { V}_{\beta_{L j}}(w_j; k_{L j}) dw_j
\Big)
\nonumber\\
&&
 \prod_{i=1}^{N_o} {\cal V}_{\alpha_i}(e^{i\phi_i};k_i)de^{i\phi_i}
|\lambda,\{q,w,n\}\rangle
\nonumber\\
&&
\langle 0|
T\Big(
\,\prod_{l=1}^{N_c} 
  c_R(k_{L l},k_{R l};Sp)\, { V}_{\beta_{R l}}({\bar w_l}; k_{R l})
 {d\bar w_l}
\Big)
| \lambda,\{-q,-w,-n\}\rangle~~
\nonumber\\
~~
\end{eqnarray}
where in the last line we have used the transposition rule
\footnote{We define the transposition of an amplitude as
\begin{eqnarray*}
&&\left[\langle0| V(z_1; k_1)\dots V(z_N; k_N) |0\rangle \right]^T
=
\left[\langle0| V(z_1; k_1)\dots V(z_N; k_N) |0\rangle \right]^{\dagger
  *}
\nonumber\\
&&=
\left[\langle0| 
V(\frac{1}{\bar z_N}; -k_N) \left(\frac{1}{\bar z_N}\right)^{2  \Delta_N}
\dots 
V(\frac{1}{\bar z_1}; -k_1) \left(\frac{1}{\bar z_1}\right)^{2  \Delta_1}
|0\rangle \right]^{
  *}
\nonumber\\
&&=
\langle0| 
V(\frac{1}{ z_N }; k_N) \left(\frac{1}{ z_N}\right)^{2  \Delta_N}
\dots 
V(\frac{1}{z_1}; k_1) \left(\frac{1}{z_1}\right)^{2  \Delta_1}
|0\rangle
\end{eqnarray*}
Notice in particular that $p^*=-p$ as it results from X reality.}
\begin{equation}
\left[ c_R(k_{L},k_{R };-Sp)\, { V}_{\beta_{R }}(\frac{1}{\bar w};
  k_{R})\right]^T= V_{\beta_R}({\bar w};k_{R}) c_R(k_{L},k_{R };Sp)
{\bar w}^2
\end{equation}
and we have changed the overall phase $\alpha_0\rightarrow\alpha_1$ due to the
reordering of cocycles.

Now the last line is invariant under the substitution
\begin{equation}
(k_R,X_R%({\bar w})
\equiv SX_R^{(NN)}
%({\bar w})
,\lambda,q,w,n) 
\rightarrow
(k_R,X_R^{(NN)}({\bar w}),S\lambda,Sq,Sw,Sn) ,
\end{equation}
so we can perform this substitution and then the renaming  
\begin{equation}
(X_L,X_R^{(NN)},\lambda,q,w,n) 
\rightarrow
(X_L^{(c)},\tilde X_R^{(c)},\tilde\lambda,\tilde q,\tilde w,\tilde n)
\end{equation}
Finally  we can rewrite eq. (\ref{corr-wc2}) as follows
\begin{eqnarray}
\label{corr-wc3}
A(N_o, N_c)
&&= 
e^{i\alpha_1}
%e^{i y_0 \left( \sum_{j=1}^{N_c}(k_{L j} -S k_{R j}) 
%               \YZS  \right)} 
%\nonumber\\
\langle\tilde 0 | \langle 0|
T \Big(
\,\prod_{j=1}^{N_c} 
  c_L(k_j;p)\, 
  c_R(k_j;\tilde p)\, 
{ V}_{\beta_{L j}}(w_j; k_{L j}) 
{\tilde V}_{\beta_{L j}}(\bar w_j; k_{R j}) d^2w_j
\Big)
\nonumber\\
&&
\prod_{i=1}^{N_o} \theta(\phi_{i}- \phi_{i+1})
 {\cal V}_{\alpha_i}(e^{i\phi_i};k_i)de^{i\phi_i}
\nonumber\\
&&
\sum_{\lambda,q,w,n}
e^{i y_0 \left( \sum_{j=1}^{N_c}(k_{L j} -S k_{R j}) 
                \YZS \right)} 
|\lambda,\{q,w,n\}\rangle
|S\tilde {\lambda},\{-S\tilde {q},-S \tilde w,- S{\tilde n}\}\rangle
\nonumber\\
~~
\label{last-A}
\end{eqnarray}
This expression involves two different sets of operators, like $a$ and
$\tilde a$ which can be interpreted as the left and right moving
operators of a closed string; because of this interpretation
 ${\cal V}_{\alpha}$ is now expressed using left moving closed string
$X_L^{(c)}(z)$ only.

As it is evident from eq. (\ref{corr-wc3}) this open string derived formalism
treats in a uniform way  compact and non compact directions, this is not
what happens in closed string formalism where  zero modes sectors
differ in compact and non compact directions.
It is however not difficult to see that the two formalisms actually yield the
same answer in zero modes sector only when the NN case is treated as a
decompactification limit as stated before.

%NORMALIZATION OF VACUUM CLOSED STRING? CONVENTIONS

\subsection{The boundary state.}
In order to derive our final expression for the boundary state
we must now take care of the last three ingredients entering the
complete amplitude given in
eq. (\ref{Ademampl}) which were left out in computing the correlator
in eq. (\ref{corr-z}): the
normalization factors, the integration region and the projective volume.
The latter is very easily taken care of 
by fixing a closed string emission vertex at 
$z_1=i e^\epsilon \rightarrow w_1=\infty$
($\epsilon\rightarrow 0$) and an open one at 
$x_{N_o}=+\infty \rightarrow \phi_{N_o}=0$, this means
that in this limit the conformal group volume can be written as 
\begin{eqnarray}
\label{Gau_Fix_used}
d V_{a b c}
&=&\frac{d^2z_1\, d x_{N_o}}{i (z_1-\bar z_1)(x_{N_o}-z_1)(x_{N_o}-\bar z_1)}
=\frac{d^2w_1\, d(e^{i\phi_{N_o}}) }{i (|w_1|^2-1) (w_1- e^{i\phi_{N_o}}) 
(e^{i\phi_{N_o}} \bar  w_1 -1)}
\nonumber\\
&\longrightarrow_{w_1\rightarrow\infty} &
\frac{d^2w_1\, d(e^{i\phi_{N_o}}) }{i e^{i\phi_{N_o}} |w_1|^4}
;
\end{eqnarray}
since the amplitude must be invariant under the gauge fixing it
follows that the correlator (when integrated over the remaining
variables) is $A(N_0,N_c)\propto e^{-i\phi_{N_o}}$ 
and hence we can decide
to only fix the closed string vertex at $z_1=i e^\epsilon$ and  let the open
string vertex be free if we integrate the so partially gauge fixed amplitude
over $\int_0^{2\pi} \frac{d e^{i\phi_{N_o}}}{2 \pi}$ and, at the same
time, we transform the $\theta(\phi_i-\phi_{i+1})$ into a periodic
$\theta_P(\phi_i-\phi_{i+1})$ to ensure only the proper sequence of
the vertices.
\COMMENTO{%%%%%%%%%%%%%%%%%%%%%%%%%%%%%%%%%%%%%%%%%%%
\footnote{
by fixing a closed string
emission vertex at $z_1=i e^\epsilon$ ($\epsilon\rightarrow0$), 
this does not fix completely the $SL(2,R)$
invariance in fact the surviving subgroup has finite volume $dV_{a b c}
\rightarrow 2\pi e^{-\epsilon}$ ( being almost the rotations group around 
$w_1(i e^\epsilon) =-ctgh(\frac{\epsilon}{2})$). The same result can
be obtained by fixing an open string state, f.x. $\phi_1=-\pi$ and
then noticing that the final result (in the closed channel) is
invariant under $\sigma$ translation so that we can let $\phi_1$ move
by introducing an integration $\int_0^{2\pi} \frac{d\phi_1}{2\pi}$.
The fixing of conformal invariance implies also
that the factor $\frac{d^2w_1}{\bar w^2_1}$ in eq. (\ref{corr-w}) is 
absent and replaced by $w_1^2$ leaving a necessary factor 
$w_1^2 {\bar  w}_1^2=|w_1|^4$ in eq. (\ref{corr-wc3}) for taking the
$\epsilon\rightarrow 0$ limit.
}
} % END COMMENTO %%%%%%%%%%%%%%%%%%%%%%%%%%%%%%%%%%%%

The normalization in eq. (\ref{Ademampl}) left out in
eq. (\ref{corr-z}) is 
${\cal C}_{0 (p)}\,{\cal N}_{o  (p)}^{N_o} \,{{\cal N}_c}^{N_c}$.
Since we want to interpret eq. (\ref{corr-wc3})
as a closed string amplitude it should be normalized as
%\footnote{
%DA FINIRE
%
%${\cal \widehat C}_0\,{\cal \widehat N}_c^{N_c+1}$ where $+1$ is due
%to the (superposition of) closed string states emitted as kets, i.e. 
% the boundary state given by the last two lines of
% eq. (\ref{corr-wc3}).
%$\frac{{\cal C}_{0 (p)} }{{\cal \widehat C}_0 {\cal \widehat N}_c} 
%{\cal N}_{o (p)}^{N_o}$ 
%}
${\cal N}_c^{N_c-1}$ {}
since the boundary state is not a normal closed string state, were it
a normal state we would normalize the amplitude as
${\cal \widehat C}_0 {\cal N}_c^{N_c}$ where ${\cal \widehat C}_0$ is
the closed string sphere normalization but this is not the proper
recipe to obtain the one point closed string emission from the
boundary state which reads $\langle \phi | B\rangle$ where $| \phi
\rangle$ is a generic closed string state   
\footnote{
In principle the closed string vertex normalization 
${\cal \widehat N}_c$ in open string formalism 
must not be equal to
the closed string vertex normalization ${\cal N}_c$ in closed string
 formalism, they are however equal
${\cal \widehat N}_c={\cal N}_c $
as follows from the request that product of two closed string vertices
 be equal in closed and open string formulations.
}.
It then follows that the boundary
state must be multiplied by
${\cal C}_{0 (p)} {\cal N}_{o (p)}^{N_o} {\cal N}_c= \frac{T_p}{2}
 {\cal N}_{o (p)}^{N_o} $. 
From this expression it is clear that the
insertion of $N_o$ open string states implies, as expected, only a
multiplicative factor ${\cal N}_o^{N_o}$ with respect to 
the ``bare'' boundary state.

Finally remembering the shift of the compact momenta in the partition
of unity given in eq. (\ref{spect_1}) we can write the boundary 
state describing multiple parallel interacting $Dp$ branes in generic
position  as
\begin{eqnarray}
\hspace{-2em}
|Dp,\{k_i,\alpha_i\}_{\{1\le i\le N_o\}}\rangle
&=&
e^{i \alpha(p)}
\Bigg [
{\cal N}_{o (p)}^{N_o}
\prod_{i=1}^{N_o} 
\int_0^{2\pi}
\theta_P(\phi_{i}- \phi_{i+1})de^{i\phi_i}
 {\cal V}_{\alpha_i}(X^{(c)}_L(e^{i\phi_i});k_i)
\Bigg ]
\nonumber\\
&&\times
e^{-i \sum_{i=1}^{N_o}k_{i} \cdot \left( \frac{x_L-S x_R}{2} \YZ \right) }
|D p\rangle
\nonumber\\
&=&
e^{i \alpha(p)}
\Bigg [
2\pi i
{\cal N}_{o (p)}^{N_o}
\prod_{i=1}^{N_o-1} 
\int_0^{2\pi} 
\theta(\phi_{i}- \phi_{i+1})de^{i\phi_i}
 {\cal V}_{\alpha_i}(X^{(c)}_L(e^{i\phi_i});k_i)
\Bigg ]
\nonumber\\
&&\times
e^{-i \sum_{i=1}^{N_o}k_{i} \cdot \left( \frac{x_L-S x_R}{2} \YZ \right) }
|D p\rangle
%\nonumber\\
%\nonumber\\
~\label{B_alpha}
\end{eqnarray}
where the open string vertices
${\cal V}_{\alpha_i}(X^{(c)}_L(e^{i\phi_i});k_i)$ are functionally
the same open string vertex operators as in eq. (\ref{V-0}) but 
now they functionally depend not on the open string fields but on 
the left moving part of the closed string fields $X^{(c)}$: to stress
this we have explicitly added $(c)$ to the notation.

The operatorial phase is given by
\begin{equation}
e^{i \alpha(p)}
=
e^{-i \pi \alpha' p_L \cdot S p_R}
\end{equation}
with $p_L=p_R$ in non compact direction
as shown in appendix (\ref{app_deter_phase}).

In the first line of eq. (\ref{B_alpha})
we have introduced a further angle $\phi_{N_o+1}=\phi_{1}$ because we
have used
a periodic $\theta_P$ and we integrate over all the $N_o$ positions 
while in the second we have set $\phi_{N_o}=0$, used a normal $\theta$
and we integrate over the first $N_o-1$ positions.
We have also introduced the usual boundary state 
\begin{eqnarray}
|D p\rangle
&=&
%\frac{
{\cal C}_{0 (p)} %}{{\cal \widehat C}_0 {\cal \widehat N}_c}
{1\over 2\pi}
\sum_{\lambda,q,w,n}
%e^{i y_0 \left( \sum_{j=1}^{N_c}(k_{L j} -S k_{R j}) 
%                \YZS \right)} 
e^{-2i y_0 \cdot ( q, \frac{w R}{ 2 \alpha' } , \frac{n}{R})}
|\lambda, ( q, \frac{w R}{ 2 \alpha' } , \frac{n}{R})\rangle
|S\tilde {\lambda}, ( -S\tilde {q},-S \frac{\tilde w R}{ 2 \alpha' } 
                  ,-S \frac{\tilde n}{R})\rangle 
\nonumber\\
\label{B_norm}
&=&
\frac{T_p}{2}
e^{-\sum_{n=1}^\infty a^\dagger_n \cdot S \cdot \tilde a^\dagger_n}|Dp_0\rangle
\end{eqnarray}
with the zero modes part
\begin{eqnarray}
|Dp_0\rangle
&=& \delta^{\ddh-p_{nc}-1}(x-y_0)|p_\alpha=0>
\sum_w e^{i \frac{wR}{\alpha'} \left( \frac{x_L-x_R}{2} -y_0\right)} |0,\tilde 0\rangle_{NN}
\sum_n e^{i \frac{n}{R} \left( \frac{x_L+x_R}{2} -y_0\right)} |0,\tilde 0\rangle_{DD}
%HOW DOES \delta(x-y_0)  COME ?
\nonumber\\
~~
\end{eqnarray}
To derive this result we have explicitly used the momentum
conservation and $S_{DD}=-1$ and $S_{NN}=+1$~\footnote{
For example to determine the dependence on $y_0$ along Neumann compact
directions, i.e. on  $y_0^n$ in eq. (\ref{B_norm}) we start from
eq. (\ref{last-A}) whose part of interest can be written as 
$\langle\tilde 0|~\langle0|~ 
e^{i y_0^n \sum_{j=1}^{N_c} (k_{L j}^n -k_{R j}^n)}
\sum_{w^n} 
| \frac{w^n R}{2\alpha'} 
+\sum_{j=1}^{N_c} k_{L j}^n 
+\frac{1}{2}\sum_{i=1}^{N_o} k_{ i}^n\rangle
~
|- \frac{\tilde w^n R}{2\alpha'} 
+\sum_{j=1}^{N_c} k_{R j}^n 
+\frac{1}{2}\sum_{i=1}^{N_o} k_{ i}^n\rangle
$ where we have used $S_{N N}=+1$ then using momentum conservation  we get
$\sum_{j=1}^{N_c} (k_{L j}^n -k_{R j}^n)= - \frac{\tilde w^n
  R}{\alpha'}  $
from which the dependence on $y_0^n$ in eq. (\ref{B_norm}) follows immediately.
The non compact Neumann case follows taking the decompactification limit.
},
%$\sum_{j=1}^{N_c}(k_{L j} -S k_{R j}) + \sum_{i=1}^{N_o} p_{i}=-2q_*$
%and that only this momentum contributes in the open channel,
moreover  we have added $DD$ and
$NN$ to remember from which open string boundary condition the pieces
were originated 
and we have supposed to have $p_{nc}=D_{NN,nc}-1= D-1-D_{DD}-D_{NN,c}$ 
non compact spacial directions along the
brane and we have defined the brane tension
$\frac{T_p}{2}
%=\frac{{\cal C}_{0 (p)} }{{\cal \widehat C}_0 {\cal \widehat N}_c}
%{1\over 2\pi}
=\frac{{\cal C}_{0 (p)} {\cal N}_c }{2\pi}
$. 
The meaning of the insertion of $exp(-i \sum_{i=1}^{N_o} k_{i}
\frac{x_L-S x_R}{2})$ in eq. (\ref{B_alpha}) (which gives a non
vanishing contribution for compact directions only) 
is to divide in the proper way the open string momentum in
the left and right moving momentum in order to ensure that 
the closed string momentum
emitted from the boundary is exactly $\sum_{i=1}^{N_o} k_{i}$.
This expression was already almost guessed in (\cite{BCD}).

The last point which must be clarified is the integration region. 
From the discussion after eq. (\ref{transf}) it is obvious that all the
moving closed string states vertices are integrated on the external
region of the unity disk $|w|>1$, this can seem unusual but the complete
amplitude associated with eq. (\ref{corr-wc3}) can nevertheless be written
as usual in the old fashion as
\begin{eqnarray}
{\cal A}(N_o, N_c)&=&
e^{i\alpha_1} 
%{\cal \widehat C}_0\,
{\cal  N}_c^{N_c-1}
\nonumber\\
\sum_{\scriptscriptstyle
      \begin{array}{l}\mbox{perm. of }\\ \{2,\dots,N_c\} 
      \end{array}}
&&
\langle \beta_{L 1},\beta_{R 1}|
{ W}^{(c)}_{\beta_{L 2},\beta_{R 2}}(1,1)
\frac{ 4\pi \Delta_c }{ \alpha' }
\dots
{ W}^{(c)}_{\beta_{L N_c},\beta_{R N_c}}(1,1)
\frac{ 4\pi \Delta_c }{ \alpha' }
|B,\{k_i,\alpha_i\}_{\{1\le i\le N_o\}} \rangle
\nonumber\\
~~~~
\end{eqnarray}
where we have introduced the closed string propagator 
$\Delta_c= \frac{\alpha'}{4\pi} \int_{|z|<1} \frac{d^2 z}{|z|^2} z^{L_0-1} {\bar z}^{\tilde L_0 -1}$
and used $W(1,1)z^{L_0} {\bar z}^{\tilde L_0 }
=z^{L_0-1} {\bar  z}^{\tilde L_0-1} W(\frac{1}{z},\frac{1}{\bar z})$ and 
$\langle \beta_{L 1},\beta_{R 1}|(L_0-1)=0$.
The somewhat strange factors $\frac{ 4\pi \Delta_c }{ \alpha' }$ are due to
the definition of the closed string propagator $\Delta_c$. 
%and the fact
Instead the factors $ {\cal N}_c$ are due to the fact in the old formalism
vertices were given without normalization factors.

\subsection{The special case of enhanced symmetry.}
Eq. (\ref{B_alpha}) is valid in the case in which branes are in a
generic position as shown in a particular case in fig.s
(\ref{figure:3open-1closed}, \ref{figure:disk-3open-1closed})  
and hence it does not need any Chan-Paton factor as
explained below eq. (\ref{Ademampl})
The case with enhanced symmetry
can be obtained when more branes are stacked on each other: in this case
we have to associate the usual Chan-Paton factor to each vertex 
as discussed below eq. (\ref{Ademampl}).
Explicitly each open string vertex acquires a Chan-Paton matrix $\Lambda$
as ${\cal V}_{\alpha}(X_L(x);k)={ V}_{\alpha}(X_L(x);k)
\rightarrow {\cal V}_{\alpha}(X_L(x);k)={ V}_{\alpha}(X_L(x);k) \Lambda$
so
that eq.(\ref{B_alpha}) becomes trivially 
\begin{eqnarray}
\label{B_alpha_CP}
|Dp,\{k_i,\alpha_i\}_{\{1\le i\le N_o\}}\rangle
&=&
e^{i \alpha(p)}~
{\cal N}_{o (p)}^{N_o}
~tr\left[
\prod_{i=1}^{N_o}
\int_0^{2\pi}
\theta_P(\phi_{i}- \phi_{i+1})de^{i\phi_j}
 {\cal V}_{\alpha_i}(X^{(c)}_L(e^{i\phi_i});k_i)
\right]~
\nonumber\\
&&
e^{-i \sum_{i=1}^{N_o} k_{i} \cdot \left( \frac{x_L-S x_R}{2} \YZ \right) }
|Dp\rangle
\end{eqnarray}

\subsection{The boundary $N_o$ Reggeon vertex.}
Eq. (\ref{B_alpha}) is the boundary state which describes from the
closed string point of view the interaction
among $N_o$ parallel branes interacting through $N_o$ open strings
whose quantum numbers are given and equal to $\alpha_i$.
We want now to write the generating function of all boundary states
with $N_o$ open string states, i.e. 
we want to write the  ``boundary'' $N_o$ Reggeon vertex.
To this purpose we introduce for any of the $N_o$ open string states
an auxiliary Hilbert space with vacuum $\langle 0_{a~i}, p_i=0|$ 
and the corresponding three Reggeon Sciuto-Della Selva-Saito 
vertex (\cite{SDS}) (see also appendix \ref{app_SDS}) so that
\begin{equation}
\langle 0_{a~ i},x_i=0|\, S_i(z)\, |\alpha_i,k_i\rangle
= V_{\alpha_i}(z;k_i)
\end{equation}
where we have defined
\begin{equation}
S_i(z)=:\exp\left(-\frac{1}{2\alpha'}\oint_{u=0, |u|<|z|} \frac{du}{2\pi i}
  X_{(c)}(u+z) \cdot \partial_u X_i(u) \right):
\end{equation}
then  the boundary (\ref{B_alpha}) can be written as
\begin{eqnarray}
\label{ReggeonB}
|Dp,\{N_o\}\rangle
&=&
e^{i \alpha(p)}~
{\cal N}_{o (p)}^{N_o}
\prod_{i=1}^{N_o} 
\oint_{|z_i|=1} d z_i
~~\theta_P(arg(z_i)- arg(z_{i+1}) )
\nonumber\\
&&
\prod_{1\le i\le N_o} <0_{a~i},x_i=0|
\prod_{1\le j< i\le N_o}
e^{-\frac{1}{2\alpha'}
    \oint_{u_i=0} \frac{d u_i}{2\pi i}
%    \oint_{u_j=0, |u_j-u_i|<|z_i-z_j|} \frac{d u_j}{2\pi i}
    \oint_{u_j=0} \frac{d u_j}{2\pi i}
    \log\left(z_i-z_j+u_i-u_j\right)
 \partial X_{i}(u_i) \partial X_{j}(v_j) 
    }:
\nonumber\\
&&
:\exp\left(-\frac{1}{2\alpha'} 
 \sum_{i=0}^{N_o}\oint_{u_i=0, |u_i|<|z_i|} \frac{d u_i}{2\pi i}
  X^{(c)}_L(u_i+z_i) \cdot \partial_u X_{i}(u_i) \right):
\nonumber\\
&&
e^{-i \sum_{i=1}^{N_o} p_{i} \cdot~ 
\left( \frac{x_L-S x_R}{2} \YZ \right)}
|Dp\rangle
%\nonumber\\
~~
\end{eqnarray}
where we have explicitly performed the necessary contractions to
normal order the vertices and $p_{i}$ is the momentum operator
acting in the i.th auxiliary open string Hilbert space.
 
As it is written the previous boundary $N_o$ Reggeon vertex 
is a linear application defined on the tensor product of $N_o$  
open string dual Hilbert spaces ${\cal H}^{*}_{open}$ 
to the closed string Hilbert space ${\cal H}_{closed}$
\begin{equation}
|Dp,\{N_o\}\rangle :  {\cal H}^{*~N_o}_{open}
\rightarrow
{\cal H}_{closed}
\end{equation}
which enjoys the fundamental property of being the ``generating
function'' of all boundary states with $N_o$ open interactions,
in fact given $N_o$ open string states $|\alpha_i>$  we can compute
eq. (\ref{B_alpha})  by
\begin{equation}
%|B,\{N_o\}\rangle
\prod_{1\le i\le N_o} |\alpha_i\rangle
\mapsto
|Dp,\{\alpha_i\}_{\{1\le i \le N_o\}}\rangle 
= |Dp,\{N_o\}\rangle \prod_{1\le i\le N_o} |\alpha_i>
\end{equation}
Once again eq. (\ref{ReggeonB}) is valid in the case in which branes are in a
generic position and hence it does not need any Chan-Paton factor as
explained below eq. (\ref{Ademampl}). 

\COMMENTO{
!!!!
The case with enhanced symmetry
can be obtained when more branes are stacked on each other in this case
we have to associate a Chan-Paton factor as discussed below
eq. (\ref{Ademampl}). 

}
\subsection{Computing a mixed disk amplitude using the boundary.}
\label{1Tc1To}
As an example we can now apply this formalism 
to recover the complete one closed tachyon - one
open tachyon amplitude whose value is $(-1)$ the value given in
eq. (\ref{1Tc-1To}) because the gauge fixing in
eq. (\ref{Gau_Fix_used}) gives this extra minus sign. 
The first step is to write the boundary state as
\begin{eqnarray*}
|D p,\{1 T_o\}\rangle
&=&
{\cal N}_{o (p)}
~e^{i \alpha(p)}
~\oint_{|z|=1} d z\,
e^{ik\cdot X_L^{(c)}(z) }\,
e^{-i k \cdot \left( \frac{x_L-S x_R}{2} \YZ \right)}
|D p\rangle
\end{eqnarray*}
and the out state as
\begin{eqnarray*}
\langle-k_L,-k_R|
&=& lim_{w\rightarrow\infty} 
\langle0,\tilde 0| 
e^{-i\pi \, p_a\, {\widehat w}^a}\,
e^{ i k_L \cdot X_L(w) +i k_R \cdot X_R(\bar w) } |w|^4
\end{eqnarray*}
then we can compute the desired amplitude as
\begin{eqnarray}
{\cal A}( 1 T_c, 1 T_o)
&=&
%{\cal \widehat C}_0\,
%{\cal \widehat N}_c
\langle-k_L,-k_R | Dp,\{1 T_o\}\rangle
\nonumber\\
&=&
%{\cal \widehat C}_0\,
%{\cal \widehat N}_c
e^{i \alpha(k)}
{\cal N}_{o (p)}
\frac{T_p}{2}  
\oint_{|z|=1} d z\,
\langle-k_L,-k_R| 
\nonumber\\
&&
~~~~~
e^{ik\cdot x_L} e^{2\alpha' k\cdot p_L \ln z}
~~\delta^{\ddh-p_{nc}-1}(x-y_0)|p_\alpha=0>
\nonumber\\
&&
~~~~~
e^{-i k \cdot \left( \frac{x_L-S x_R}{2} \YZ \right)}
~\sum_w e^{i \frac{wR}{\alpha'} \left( \frac{x_L-x_R}{2} -y_0\right)} 
|0,\tilde 0\rangle_{NN}
~\sum_n e^{i \frac{n}{R} \left( \frac{x_L+x_R}{2} -y_0\right)}
|0,\tilde 0\rangle_{DD}
\nonumber\\
\end{eqnarray}
In computing the previous amplitude we must consider the non compact
Dirichlet directions as a limit of compact ones so that we can write
\begin{eqnarray}
{\cal A}( 1 T_c, 1 T_o)
&=&
%{\cal \widehat C}_0\,{\cal \widehat N}_c^2
%{\cal \widehat N}_c
e^{i \alpha(k)}
{\cal N}_{o (p)}
\frac{T_p}{2} 
\oint_{|z|=1} d z\,
\nonumber\\
&&
e^{-2\alpha' k^\nu (k_L+k)_\nu \ln z}
\left(2\pi \delta(k_L^\nu+k_R^\nu+k^\nu)\right)^{D_{NN,nc}}
\nonumber\\
&&
\lim_{R^\delta\rightarrow \infty}
e^{-2\alpha' k^\delta (k_L+k)_\delta  \ln z}
\sum_{n^\delta} \delta_{k_R^\delta - \frac{1}{2} k^\delta +
  \frac{n^\delta}{2 R^\delta},0 }
~
\delta_{k_L^\delta + \frac{1}{2} k^\delta +
  \frac{n^\delta}{2 R^\delta},0 }
e^{-i \frac{n^\delta }{R} y_0^\delta }
\nonumber\\
&&
e^{-2\alpha'k^n (k_L+k)_n  \ln z}
\sum_{w^n} \delta_{k_R^n + \frac{1}{2} k^n - \frac{w^n R}{2 \alpha'},0 }
~
\delta_{k_L^n + \frac{1}{2} k^n + \frac{w^n R}{2 \alpha'},0 }
e^{-i \frac{w^n R}{\alpha'} y_0^n }
\nonumber\\
&&
e^{-2\alpha' k^d (k_L+k)_d  \ln z}
\sum_{n^d} \delta_{k_R^d - \frac{1}{2} k^d + \frac{n^d}{2 R},0 }
~
\delta_{k_L^d + \frac{1}{2} k^d + \frac{n^d}{2 R},0 }
e^{-i \frac{n^d }{R} y_0^d }
\nonumber\\
\end{eqnarray}
where the first line is from non compact Neumann directions,
the second line from non compact Dirichlet directions,
the third line from  compact Neumann directions and
the fourth line from compact Dirichlet directions.
We have used the index $n$ ($\nu$) for (non) compact NN
directions and $d$ ($\delta$) for  (non) compact DD ones.

The Kronecker $\delta$s in non compact Dirichlet directions imply that
the closed tachyon momentum is unconstrained while the open tachyon
``momentum'' which is interpreted as a distance between two branes 
must vanish $k^\delta=0$.

It is also worth noticing as the insertion of 
$e^{-i k \cdot \left( \frac{x_L-S x_R}{2} \YZ \right)}$
in the boundary state is fundamental for finding a consistent solution
of the compact $\delta$s, explicitly for the Neumann directions we
find $n_{closed}^n+n_{open}^n+\theta^n=0$ and $w_{closed}^n+w^n=0$;
would the insertion not be present we would have found an inconsistent
(with the translation invariance)
linear system, i.e.
$\frac{1}{2}n_{closed}^n+n_{open}^n+\theta^n=0$ from the left moving
sector and $\frac{1}{2}n_{closed}^n=0$ from the right moving one.

Finally we get
\begin{eqnarray}
{\cal A}( 1 T_c, 1 T_o)
&=&
%{\cal \widehat C}_0\,{\cal \widehat N}_c^2
%{\cal \widehat N}_c
e^{i \alpha(k)}
{\cal N}_{o (p)} 
\frac{T_p}{2} 
\oint_{|z|=1} d z\,
e^{i(k_L-Sk_R  )y_0 }
e^{-2\alpha'(k_L+k) k\ln z}
\nonumber\\
&&
~~~~~
\left(2\pi \delta(k_L+k_R+k)\right)^{D_{NN,nc}}
\delta_{k_L+Sk_R+k,0}^{D_{DD}+D_{NN,c}}
%\delta_{\theta,0}
\nonumber\\
&=&
i~e^{i \alpha(k)}~
{\cal C}_{0 (p)}\, {\cal  N}_c
{\cal N}_{o  (p)}
e^{i(k_L-Sk_R)y_0 }
\left(2\pi \delta(k_L+k_R+k)\right)^{D_{NN,nc}}
\delta_{k_L+Sk_R+k,0}^{D_{DD}+D_{NN,c}}
%\delta_{\theta,0}
\nonumber\\
\end{eqnarray}
where  we have used the mass shell conditions 
$\alpha' k^2=\alpha' k_R^2=\alpha' k_L^2=1$ to write 
\begin{equation}
e^{-2\alpha'(k_L+k)  k\ln z}
= e^{-2\alpha'(\frac{k_L+S k_R}{2}+k+\frac{k_L-S k_R}{2})  k\ln z}
= e^{-\alpha' \left( k^2 -(k_L-S k_R)(k_L+S k_R) \right)\ln z}
=z^{-1} 
\end{equation}
 in order to perform the $z$ integration and the
relation 
%$T_p=\frac{{\cal C}_{0 (p)} }{{\cal \widehat C}_0 {\cal \widehat    N}_c}{1\over 2\pi}$
$\frac{T_p}{2} =\frac{{\cal C}_{0 (p)} {\cal N}_c }{ 2\pi}$
 from eq. (\ref{B_norm}) to rewrite the normalization 
coefficient.
Finally because of the momentum conservation 
$2 \alpha' k_L \cdot S k_R=-1$
 the phase factor is
\begin{equation}
e^{i \alpha(k)}= e^{-i \pi \alpha' k_L \cdot S k_R} = i
\end{equation}
which together the factor $i$ already present gives the required factor
$-1$.

\subsection{One loop $N$ tachyons amplitudes.}
\begin{figure}[hbtp]
\begin{minipage}[t]{0.45\linewidth}
\scalebox{0.9}{
\includegraphics[type=eps,ext=.eps,read=.eps,width=0.9\textwidth]{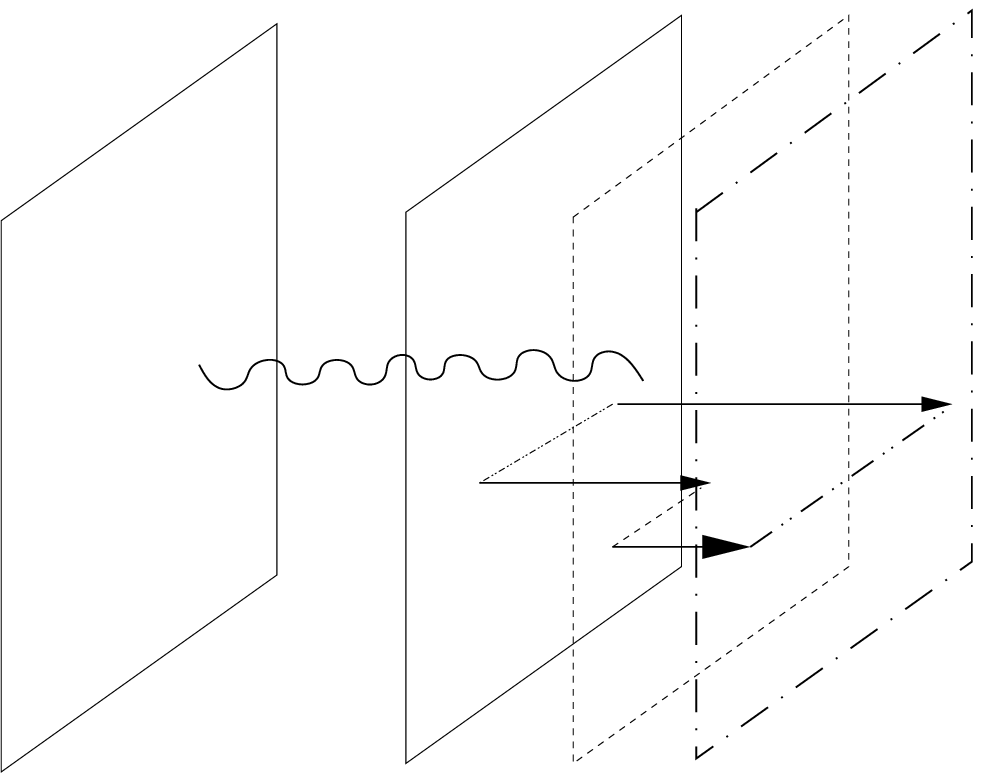}
}
 \caption{One brane is gravitationally interacting with 
3 parallel branes which interact among themselves with 3 open strings.
 }
\label{figure:3open-1}
\end{minipage}
\hskip 1cm
\begin{minipage}[t]{0.45\linewidth}
\scalebox{0.9}{
\includegraphics[type=eps,ext=.eps,read=.eps,width=0.9\textwidth]{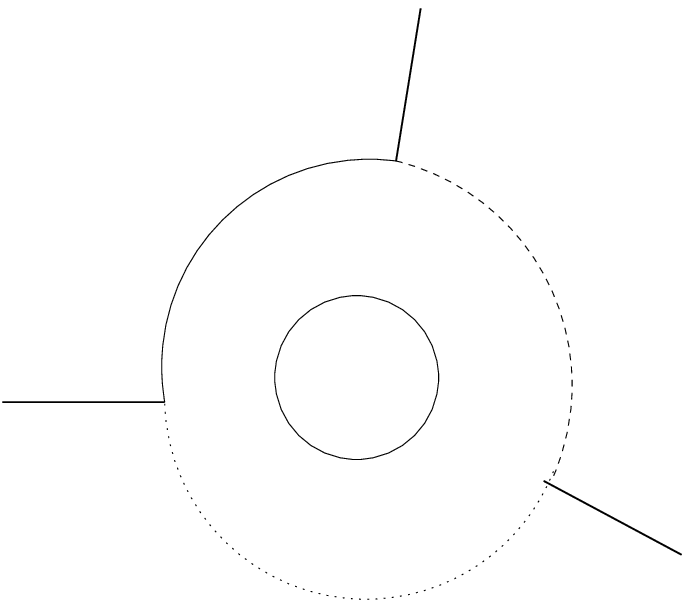}
}
 \caption{The previous figure interpreted in the open channel.
 }
\label{figure:3open-1loop}
\end{minipage}
\end{figure}

As a final example of application of the interacting boundary state we want to
compute the complete one loop planar amplitude having $N$ open string
tachyons,
normalization included.
% in the internal border and $N_\pi$ on the external in the closed
% string channel as
Explicitly we compute the gravitational interaction between one $Dp$
brane and $N$ interacting parallel $Dp$ branes in a generic position
as it is depicted in fig. (\ref{figure:3open-1}) in a special case,
then we reinterpret this result in the open channel as  
 in fig. (\ref{figure:3open-1loop}).
The closed string computation is given by 
\begin{eqnarray}
{\cal A}_{N~tach, 1~loop}
&=&
%\hat {\cal C}_0 \widehat {\cal N}_c^2
\langle Dp | 
\Delta_c 
|Dp,\{N\}\rangle \prod_{1\le i\le N} |k_i\rangle
\nonumber\\
&=&%\hat {\cal C}_0 \widehat {\cal N}_c^2
\langle Dp | 
\Delta_c 
| Dp, \{ k_i\}_{\{ 1\le i \le N\}}\rangle
\nonumber\\
&=& %\hat {\cal C}_0 \widehat {\cal N}_c^2 
{\cal N}^N_{o (p)}
\left(\frac{T_p}{2} \right)^2
~\langle Dp | ~e^{-i \alpha(p)}
~\frac{\alpha' \pi}{2}  \int_0^\infty d\tau e^{-\pi \tau(L_0 +\tilde L_0-2)}
\delta_{L_0 ,\tilde L_0}
\nonumber\\
&&
e^{i \alpha(p)}~
2\pi i
\int_0^{2\pi}
\prod_{i=1}^{N-1}  \theta(\phi_{i}- \phi_{i+1})de^{i\phi_i}
\prod_{i<j} ( e^{i\phi_i} - e^{i\phi_j} )^{2\alpha' k_i\cdot k_j}
\nonumber\\
&&
e^{i \sum_i k_{i} \cdot X^{(c)(-)}_L(e^{i\phi_i}) }
e^{i \sum_j k_{j} \cdot X^{(c)(+)}_L(e^{i\phi_j}) }
e^{-i \sum_i k_{i} \cdot \left( \frac{x_L-S x_R}{2} \YZ \right) }
|Dp\rangle
\end{eqnarray}
where $\phi_{N}=0$ and 
$|Dp\rangle$ is the plain boundary state associated with 
a $Dp$ brane with boundary conditions dictated by the reflection
matrix $S$.
A straightforward computation gives
\begin{eqnarray}
&&{\cal A}_{N~tach, 1~loop}=
\nonumber\\
&=& (2\pi)^2 i 
%~\hat {\cal C}_0 \widehat {\cal N}_c^2 
\frac{\alpha'}{4}
{\cal N}^N_{o (p)} 
\left(\frac{T_p}{2} \right)^2
\int_0^\infty d\tau
\int_0^{2\pi} \prod_{i=1}^{N-1}  de^{i\phi_i} ~\theta(\phi_{i}- \phi_{i+1})
\nonumber\\
&&
\left[ 2\pi \delta(\sum_i k_{i}^\nu )\right]^{D_{N N,n c}}
\times
 \delta_{\sum_i k_{i}^\delta }^{D_{D D,n c}}
\left( \frac{2}{\alpha' \tau}\right)^{D_{D D,n c}/2}
e^{- \frac{2}{\alpha' \tau} 
\left( y_\pi^\delta -y_0^\delta +\alpha' \sum_i k_{i}^\delta\phi_i\right)^2 }
\nonumber\\
&&
\delta_{\sum_i k_{i}^n }^{D_{N N,c}}
\sum_w e^{ - \frac{1}{2} \pi \alpha' \tau \left(\frac{w^n R}{\alpha'} \right)^2}
e^{i \frac{w R}{\alpha'} 
     \left(   y_\pi^n -y_0^n +\alpha' \sum_i k_{i}^n\phi_i\right) 
  }
\times
\delta_{\sum_i k_{i}^d }^{D_{D D,c}}
\sum_n e^{ - \frac{1}{2} \pi \alpha' \tau \left(\frac{n }{R} \right)^2}
e^{i \frac{n}{R} 
     \left(   y_\pi^d -y_0^d +\alpha' \sum_i k_{i}^d\phi_i \right) 
  }
\nonumber\\
&&
\left( \prod_{n=1}^\infty \frac{1}{1-e^{-2\pi \tau n} } \right)^{D-2}
\times
\prod_{i<j} \left( \prod_{m=1}^\infty
\frac{ 1- 2 \cos \phi_{ij}~  e^{-2\pi \tau m } + e^{-4\pi \tau m } }
     { (1-e^{-2\pi \tau m })^2 }
\right)^{ 2\alpha' k_i\cdot k_j}
 ( e^{i\phi_i} - e^{i\phi_j} )^{2\alpha' k_i\cdot k_j}
\nonumber\\
~
\label{NTach1Loop}
\end{eqnarray}
where the first two terms in the last line are due to non zero modes
(the ghosts account for the $-2$ in the exponent $D-2$) with
$\phi_{i j}=\phi_i-\phi_j$,
in last but one line we have the compact zero modes contributions
and in the third line there are the contributions from non compact
zero modes: the non compact contribution from DD directions 
must again be understood as the
decompactification limit since we otherwise would miss the
conservation of distances which is a Kronecker delta and not a Dirac
delta.
%Moreover we have not explicitly specified which directions we are
%summing considering in expression like $ \sum_i  k_{i}$ since this can
%be deduced from the corresponding $\delta$  which 
%explicitly shows which kind of direction we are considering.
As before we have used the index $n$ ($\nu$) for (non) compact NN
directions and $d$ ($\delta$) for  (non) compact DD ones.

In eq. (\ref{NTach1Loop}) the contribution from non zero modes was
easily computed using 
\begin{equation}
\langle0,\tilde 0| e^{-s_0 a \tilde a} e^{a^\dagger A} e^{a B}
e^{-s_\pi a^\dagger \tilde a^\dagger} | 0,\tilde 0\rangle=
\frac{1}{1-s_0 s_\pi} e^{ \frac{A B s_0 s_\pi}{1-s_0 s_\pi} }
\end{equation}

The string amplitude (\ref{NTach1Loop}) 
for the special case of $D25$ ($S=\uno$) with
all non compact directions reduces
to the well know 1 loop $N$ tachyons  open string amplitude:
\begin{eqnarray}
&&{\cal A}_{N~tach, 1~loop}=
\nonumber\\
&=&  
%\frac{\alpha'}{4}
{\cal C}_{1 (25)}
{\cal N}^N_{o (25)} 
(2\pi)^{26} \delta^{26}(\sum_i k_{i}^\nu )
~
\int_0^\infty d\tau
\frac{2\pi}{2}
\int_0^{2\pi} \prod_{i=1}^{N-1}  
\frac{d\phi_i}{2} ~\theta(\phi_{i}- \phi_{i+1})
\nonumber\\
&&
\left( \prod_{n=1}^\infty \frac{1}{1-e^{-2\pi \tau n} } \right)^{D-2}
\times
\prod_{i<j} \left( \sin \frac{\phi_{i j}}{2} \prod_{m=1}^\infty
\frac{ 1- 2 \cos \phi_{ij}~  e^{-2\pi \tau m } + e^{-4\pi \tau m } }
     { (1-e^{-2\pi \tau m })^2 }
\right)^{ 2\alpha' k_i\cdot k_j}
\nonumber\\
~
\end{eqnarray}
upon the use of
%$2\pi  \hat {\cal C}_0 \widehat {\cal N}_c^2 ~T_{25}^2=1$
%$2\pi  \hat {\cal C}_0 \widehat {\cal N}_c^2 ~T_{25}^2=1$
%(which is equivalent to ${\cal C}_{0 (25)}^2=2\pi  \hat {\cal C}_0$ ),
momentum conservation, 
mass shell condition $\alpha' k_i^2=1$,  
$2\alpha' \sum_{i<j}k_i\cdot k_j= -N$
and  $\phi_{N}=0$  to write
\begin{eqnarray}
&&\prod_{k=1}^{N-1} d{e^{i\phi_k}} 
\prod_{1\le i<j\le N} ( e^{i\phi_i} - e^{i\phi_j} )^{2\alpha' k_i\cdot k_j}
\nonumber\\
&=&
\prod_{k=1}^{N-1} d{e^{i\phi_k}}
\prod_{i<j} \left[ e^{i\frac{\phi_i+\phi_j}{2} 2\alpha' k_i\cdot k_j}
(2i~\sin \frac{\phi_{i j}}{2} )^{2\alpha' k_i\cdot k_j} \right]
\nonumber\\
&=&
\prod_{k=1}^{N-1}~d{\phi_k}
\prod_{i<j} (2 i~\sin \frac{\phi_{i j}}{2} )^{2\alpha' k_i\cdot k_j}
=
\frac{1}{2 i}
\prod_{k=1}^{M-1} \frac{d{\phi_k}}{2}
\prod_{i<j} (\sin \frac{\phi_{i j}}{2} )^{2\alpha' k_i\cdot k_j}
\end{eqnarray}

%%%%%%%%%%%%%%%%%%%%%%%%%%%%%%%%%%%%%%%%%%%%%%%%%%%%%%%%%%%%%%%%%%%%%%%%%%%%%%
%%%%%%%%%%%%%%%%%%%%%%%%%%%%%%%%%%%%%%%%%%%%%%%%%%%%%%%%%%%%%%%%%%%%%%%%%%%%%%
%%%%%%%%%%%%%%%%%%%%%%%%%%%%%%%%%%%%%%%%%%%%%%%%%%%%%%%%%%%%%%%%%%%%%%%%%%%%%%
%%%%%%%%%%%%%%%%%%%%%%%%%%%%%%%%%%%%%%%%%%%%%%%%%%%%%%%%%%%%%%%%%%%%%%%%%%%%%%
\section{Conclusions.}
\label{par-conc}
In this article we have shown how to compute boundary states which
describe $N$ parallel branes in generic position, i.e. not
superimposed, interacting through $N$ open string states in
eq. (\ref{B_alpha}). We have also computed the corresponding boundary
states when the branes are superimposed as in eq.(\ref{B_alpha_CP}).

The result is not completely unexpected but there are some points which
are not trivial:
\begin{itemize}
\item
when acting on the trivial boundary state with open string vertex
operators we have to substitute $X_{open}$ with the {\sl left} moving
part of the closed string $X_L$ in order to obtain a T-duality
invariant formulation;
\item
the non compact Dirichlet directions must be treated as a
decompactification limit;
\item
in compact directions we have to equally divide the momenta coming
from open string vertices between left and right closed string
momenta, as it is explicit in eq. (\ref{B_alpha}).
\end{itemize}    

Another point which is worth noticing is that this derivation is
valid for the bosonic free string but this could be in principle extended to
other CFTs since the fundamental starting point is the possibility of
writing the closed string vertices as a product of two open string
ones. This possibility is quite natural when one starts from the open
string $\sigma$ model as discussed around eq. (\ref{sigma_model}):
in particular this is clear for the bosonic/ NS-NS gravitational
sector which couples
universally in both bosonic and fermionic strings but it is also true for
the fermionic and RR sectors where branes couple only to a subsector of
all the possible closed string states.
 
%%%%%%%%%%%%%%%%%%%%%%%%%%%%%%%%%%%%%%%%%%%%%%%%%%%%%%%%%%%%%%%%%%%%%%%%%%%%%%
%%%%%%%%%%%%%%%%%%%%%%%%%%%%%%%%%%%%%%%%%%%%%%%%%%%%%%%%%%%%%%%%%%%%%%%%%%%%%%

\vskip 0.5cm
\noindent
{\bf Acknowledgments}

\noindent
The author thanks M. Bill\'o and A. Lerda for discussions and the 
Niels Bohr Institute for the hospitality during the completion of this paper.
This work supported in part by the European Community's Human Potential
Program under contract HPRN-CT-2000-00131 Quantum Spacetime.
It is also partially supported by the Italian MIUR under the program
``Teoria dei Campi, Superstringhe e Gravit\'a''.

\appendix
\section{Conventions.}
\label{conventions}
\begin{itemize}
\item Indices:\\
 $\alpha=0\dots \ddh-1$ for non compact directions
and $a=\ddh\dots D-1$ for compact ones;
$n$ ($\nu$) for (non) compact NN
directions and $d$ ($\delta$) for  (non) compact DD ones.
\item Amplitudes normalizations.\\
If the $d$ dimensional YM action is given by
$S_{YM}=\int d^d x ~tr(F_{\mu\nu} F^{\mu\nu})$ with 
$F_{\mu\nu}=\partial_\mu A_\nu -\partial_\nu A_\mu 
-i g_d [A_\mu,A_\nu]$,
$A_\mu=A_\mu^a \lambda^a$ and 
$tr(\lambda^a\lambda^b)=\frac{1}{2}\delta^{a b}$,
$tr(\lambda^a[\lambda^b,\lambda^c])=\frac{i}{2}f^{a b c}$, 
it is possible to derive from direct comparison with field theory
as in ref. \cite{1loop} 
$$
{\cal N}_{o (d-1)}=g_d \sqrt{2\alpha'}
\,\, , \,\,
{\cal C}_{0 (d-1)}=\frac{1}{(g_d \alpha')^2}
\,\, , \,\,
\frac{ {\cal C}_{h+1 (d-1)} }{ {\cal C}_{h (d-1)} }
=\frac{(g_d \alpha')^2}{(2 \pi)^d (2
  \alpha')^{d/2} } 
$$
and from unitarity 
$$
{\cal C}_{0 (d-1)}{\cal N}_{o (d-1)}^2 \alpha'=2
$$
when using $\Delta_o=\int_0^1 dx~x^{L_0-2} $ 
($L_0= \alpha' p^2 +\sum_1^\infty n a^\dagger_n \cdot a_n$).
In a similar way it follows from closed string unitarity
$$
{\cal \widehat C}_0{\cal \widehat N}_c^2 \alpha'=4 \pi
$$
when using as closed string propagator 
$\Delta_c= \frac{\alpha'}{4 \pi}\int_{|z|<1} \frac{d^2 z}{|z|^2} z^{L_0-1} {\bar z}^{\tilde
  L_0 -1}$
($L_0= \alpha' p_L^2 +\sum_1^\infty n a^\dagger_n \cdot a_n$,
$\tilde L_0= \alpha' p_R^2 +\sum_1^\infty n \tilde a^\dagger_n \cdot
\tilde a_n$,
).

The vertex normalization factors ${\cal N}_{o (d-1)}$, ${\cal N}_c$ and ${\cal
  \widehat N}_c$ are common to all the states but depend on the number
  of compact and non compact directions.
\item Vacua normalizations in non compact directions.\\
For each non compact direction of open string with NN b.c. or of
closed string: 
      $\langle k| k'\rangle= 2\pi\delta(k-k')$ with $\langle k| =|k\rangle^\dagger$,
      $|x=0\rangle\equiv \delta(x)|p=0\rangle$,

For each non compact direction with DD b.c. since only $|0\rangle$ has
      finite energy: 
      $\langle 0| 0\rangle= 1$
but the spectral decomposition of the unity is
$\uno=\lim_{R\rightarrow\infty}\sum_w |w\rangle\langle w|$;
For each compact direction $\langle n| m \rangle=\delta_{n,m}$.
\item 
$z\in\complessi \Leftrightarrow -\pi<arg(z)\le\pi$,
      $z\in\Hplus \Leftrightarrow 0\le arg(z)\le\pi$
\item
Open string modes expansions:
$$
X(z,\bar z)=\frac{1}{2}\left(X_L(z)+X_R(\bar z)\right)
$$
$$
X^{\mu}_{L/R}(z) = 
q^{\mu}\pm y^\mu_0 
-{\rm i}\, \sqrt{2\alpha'}\, a_0^{\mu}{}\ln z +
{\rm i}\,\sqrt{2\alpha'}\,
\sum_{n\not =0} {sgn(n)\over {\sqrt {|n|}}} a_n^{\mu}z^{-n}
$$
where $z=e^{\tau_E+i\sigma}\in \Hplus$ ($\tau_E$ is the Wick rotated
time) and 
the logarithm entering the string expansion is defined to have a
cut at $arg(z)=-\pi$, i.e. $-\pi<arg(z)\le +\pi$
moreover $log(\bar z)\equiv \overline{log(z)}$
and $a^\mu_0=\sqrt{2\alpha'}p^\mu$ in the NN case.

The commutation relations are ( $a_{-n}=a_n^\dagger$)
$$
[q^\mu, p^\nu]= i \eta^{\mu \nu}
~~~~
[a_m^\mu, a^{\dagger \nu}_n]= \delta_{m,n} \eta^{\mu \nu}~~
m,n>0
$$

The spectrum of the momentum operator is
$$
k=\left\{
\begin{array}{ll} 
k^\alpha &  \textrm{non compact NN bc}\\
\frac{n^n +\theta_0^n-\theta_\pi^n}{R} 
&  \textrm{compact NN bc}\\
\frac{y_\pi^\delta - y_0^\delta}{2\pi\alpha'} & \textrm{non compact DD bc}\\
\frac{w^d ~ R}{\alpha'}+\frac{y_\pi^d - y_0^d}{2\pi\alpha'} & \textrm{compact DD bc}
\end{array} 
\right.
$$
where $y_0$ ($y_\pi$) is the position of the brane at $\sigma=0$
($\sigma=\pi$) and $2\pi \theta_0^n$ ($2\pi \theta_\pi^n$)
is the Wilson line on the brane at $\sigma=0$ ($\sigma=\pi$). 
\item
The closed string modes expansion is given in analogous manner by
$$
X(z,\bar z)=\frac{1}{2}\left(X^{(c)}_L(z)+{\tilde X^{(c)}_R}(\bar
z)\right)
$$ 
where now $z=e^{2\tau_E+2i\sigma}\in \complessi$, $X^{(c)}_L$ has the same
expansion as for the open string but with $y_0=0$  and ${\tilde X}_R^{(c)}$
expands as $X_L^{(c)}$ but with tilded operators, the commutation
relations read
$$
[q_L^\mu, p_L^\nu]=[q_R^\mu, p_R^\nu]= i \eta^{\mu \nu}
~~~~
[a_{L m}^\mu, a^{\dagger \nu}_{L n}]= 
[\tilde a_{R m}^\mu, \tilde a^{\dagger \nu}_{R n}]=\delta_{m,n} \eta^{\mu \nu}~~
m,n>0
$$
In the non compact case we identify $x_L=x_R=x$ and
$p_L=p_R=\frac{p}{2}$ where $p$ has continuum spectrum while in
compact directions $p_L$ ($p_R$) has discrete spectrum given by
$k_L=\frac{1}{2}\left(\frac{n}{R}+ \frac{w R}{\alpha'} \right)$
($k_R=\frac{1}{2}\left(\frac{n}{R}- \frac{w R}{\alpha'} \right)$).

\end{itemize}

%%%%%%%%%%%%%%%%%%%%%%%%%%%%%%%%%%%%%%%%%%%%%%%%%%%%%%%%%%%%%%%%%%%%%%
%%%%%%%%%%%%%%%%%%%%%%%%%%%%%%%%%%%%%%%%%%%%%%%%%%%%%%%%%%%%%%%%%%%%%%

\section{Phases and Analytic continuations.}
\label{app_SDS}
We start writing the three Reggeon Sciuto-Della Selva-Saito vertex
(\cite{SDS}) for a single coordinate
\begin{equation}
S(z)=:\exp\left(-\frac{1}{2\alpha'}\oint_{u=0, |u|<|z|} \frac{du}{2\pi i}
  X_{aux}(u+z) \partial_u X(u) \right):
\end{equation}
where the normal ordering is taken with respect to both the auxiliary
$X_{aux}$ and the usual $X(u)\equiv X_L(u)$.
Performing the integral over $u$ 
the previous equation becomes ($\sqrt 0 \equiv 1$)
\footnote{
In the case of a non compact direction the left and right moving parts
cannot be factorized because of the zero modes therefore the complete
vertex reads:
$$
S(z,{\bar z})=:\exp\left(\frac{i}{\sqrt{2\alpha'}}
\left[a_0 X_{aux}(z,{\bar z})
     +\sum_{n=1}^\infty \frac{\sqrt{n}}{n!} 
       (a_n\partial^n+\tilde a_n {\bar\partial}^n) X_{aux}(z,{\bar z})
\right]
     \right): 
$$
}
\begin{equation}
S(z)=:\exp\left(\frac{i}{\sqrt{2\alpha'}}
      \sum_{n=0}^\infty \frac{\sqrt{n}\,a_n}{n!} \partial^nX_{aux}(z)
     \right): 
\end{equation}  
This vertex can be thought of as a generating functional of all
vertices\footnote{
This yields the proper expression for the open string vertex only for
the emission from $\sigma=0$ when $z=x$. 
but the right expression of the SDS vertex
for emission from the $\sigma=\pi$ border is
$$
 S_\pi(\zeta)=:S(\zeta) \exp\left(-i\pi\,2\alpha'\, p_0 p_{0(aux)}\right):
$$
The necessity of the cocycle follows from the obvious request of
commutativity of the product of a vertex for the emission from
$\sigma=0$ border with one from $\sigma=\pi$ border.
}
:
\begin{equation}
\langle 0_a,x=0|\, S(z)\, |\alpha,k\rangle= V_\alpha^{(aux)}(z;k)
\end{equation}

We use this formalism in order to exam in general the phases occurring
when performing the analytic continuation of the product of two such
vertices.
We first compute the product of two SDS vertices
\begin{eqnarray}
\label{2_SDS_prod}
S(z)\, S(w)\,\, &=&\, 
:S(z)\, S(w)\, 
\nonumber\\
&&e^{-\frac{1}{2\alpha'}
    \oint_{u=0} \frac{du}{2\pi i}
    \oint_{v=0, |v-u|<|z-w|} \frac{dv}{2\pi i}
    \log\left(z-w+u-v\right)
 \partial X_{(1)}(u) \partial X_{(2)}(v) 
    }:
\nonumber\\
&=&\, 
:S(z)\, S(w)\, e^{\log(z-w)\, 2\alpha'\,p^{(1)}p^{(2)} }...: 
\,\,\,\,
|z|>|w|
\nonumber\\
\end{eqnarray}
in this expression the $log$ has to be interpreted as a shorthand version of
$$
\log\left(z-w+u-v\right)\equiv
\log(z)
%+\log\left(1-\frac{w}{z}\right)
%+\log(1+\frac{u-w}{z-w})
-\sum_{n=1}^\infty \frac{1}{n} \left(\frac{w}{z}\right)^n
-\sum_{n=1}^\infty \frac{1}{n} \left(\frac{v-u}{z-w}\right)^n
$$ 
%where the two last $log$ are actually given by a series.
and we have made use of the well known expression
\begin{equation}
X(z)X(w)=:X(z)X(w):-2\alpha'\, \log(z-w)
\,\,\,\, |z|>|w|
\end{equation}
Then we compare this equation (\ref{2_SDS_prod}) with the analogous product
$S(w)\,S(z)$ for $|z|<|w|$ and we use
\begin{equation}
\left[ \log (z-w )\right]_{an. ~cont.} 
= 
\log (w-z) + i\pi~ sign(arg(z)-arg(w))
~~~~
-\pi< arg(z),arg(w)<\pi
\end{equation}
we get the desired phase
\footnote{
In a similar way we deduce that in the non compact case the vertices commute
$$
\left[S(z,{\bar z})\, S(w,{\bar w})\right]_{an.cont}
=  S(w,{\bar w})\,S(z,{\bar z}) 
$$
because of 
$ X(z,\bar z)X(w,\bar w)=:X(z,\bar z)X(w,\bar w):-\alpha'\, \log |z-w|$
}
\begin{equation}
\label{phase_normal}
\left[S(z)\, S(w)\right]_{an.cont}
=  S(w)\,S(z) 
e^{
 +i\pi\,sgn(arg(z)-arg(w))\,2\alpha'\, p_{0(1)}p_{0(2)}
}
\end{equation}
with $-\pi<arg(z),arg(w)\le+\pi$.
In the generic case the phase of the previous equation becomes
\begin{equation}
\label{phase_general}
\pi\,sgn(arg(z)-arg(w)) \Rightarrow
2\pi(n_z-n_w)+\pi\, sgn([\phi_z]-[\phi_w])
\end{equation} 
where we have defined $arg(z)=[\phi_z]+2\pi n_z$ with
$-\pi<[\phi_z]\le \pi$ and $n_z\in\interi$.

%%%%%%%%%%%%%%%%%%%%%%%%%%%%%%%%%%%%%%%%%%%%%%%%%%%%%%%%%%%%%%%%%%%%%%
%%%%%%%%%%%%%%%%%%%%%%%%%%%%%%%%%%%%%%%%%%%%%%%%%%%%%%%%%%%%%%%%%%%%%%

\section{Details on the cocycles computations.}
\subsection{Closed string case.}
\label{details_Cstr}
In this section we give the details on the derivation of the closed
string cocycles.
In section (\ref{clos_stri_coc_sub}) we introduced the phase 
$e^{i\Phi_{(c)}(k_1,k_2)}$ as the phase which arises when computing
the normal ordering of the product of two closed string vertices $\cal
W$, 
explicitly
\begin{eqnarray}
{\cal W}^{(c)}_{\beta_L,\beta_R}(z,\bar z; k_1)
{\cal W}^{(c)}_{\alpha_L,\alpha_R}(w,\bar w; k_2)
&=&
e^{i\Phi_{(c)}(k_1,k_2)}
c(k_{L1}+k_{L2},k_{R1}+k_{R2}; p_L,p_R)*
\nonumber\\
&&
V_{\beta_L}(z;k_{L1})
V_{\alpha_L}(w;k_{L2})
\,\, \, \,
%\nonumber\\
%&&
{\tilde V}_{\beta_R}({\bar z};k_{R1})
{\tilde V}_{\alpha_R}({\bar w};k_{R2})
\nonumber\\
\label{OPE_phase_closed_app0}
\end{eqnarray}
with
\begin{eqnarray}
\Phi_{(c)}(k_1,k_2)
&=&
-\pi\, \alpha'\left[
\left( \bb\, k_{L2} +\cc \, k_{R2}\right) \cdot k_{L1} 
+\left( \dd\, k_{L2} +\ee \, k_{R2}\right) \cdot k_{R1} 
\right]
\label{OPE_phase_closed_app}
\end{eqnarray}
where the matrices $\bb, \cc, \dd$ and $\ee$ may only have non
vanishing  entries in compact directions, i.e. for example $\bb^a_b$.
  
With the help of the well known formula 
($-\pi\le arg(z),arg(w)\le+\pi$, $|z|<|w|$) (see also app. \ref{app_SDS})
$$
\left[V(z;k_1)\,V(w;k_2)\right]_{\mbox{an.con.}}
= V(w;k_2)\,V(z;k_1)
e^{+i\pi\,sgn(arg(z)-arg(w))\,2\alpha'\, k_1 \cdot k_2}
$$
and by comparing with the analogous product of the vertices 
in the opposite order
$$ 
{\cal W}^{(c)}_{\alpha_L,\alpha_R}(w,\bar w; k_2)
{\cal W}^{(c)}_{\beta_L,\beta_R}(z,\bar z; k_1)
$$
it is not difficult to see that the constraint we want to
impose on the cocycles coefficients in order to implement the commutativity is
\footnote{
Because of (\ref{phase_general}) and (\ref{k_products}) the same
constraint holds in the world-sheet minkowskian version where
$arg(z)=\tau+\sigma$. } 
\begin{eqnarray}
&&e^{i\Phi_{(c)}(k_1,k_2)-i\Phi_{(c)}(k_2,k_1)
            +i\pi sgn(arg(z)-arg(w))
              2\alpha'(k_{L1}\cdot k_{L2}-k_{R1}\cdot k_{R2})
  }=
\nonumber\\
&&
e^{-i\pi\, \alpha'
            \left[
              (\bb-\bb^T)k_{L2}\cdot k_{L1}
              +(\ee-\ee^T)k_{R2}\cdot k_{R1}
		  \right]
}
\times
\nonumber\\
&&
e^{-i\pi\, \alpha'
            \left[
            (\cc-\dd^T )k_{R2}\cdot k_{L1}
                  -(\cc-\dd^T) ^T k_{L2}\cdot k_{R1}
            -2 sgn(arg(z)-arg(w))(k_{L1}\cdot k_{L2}-k_{R1}\cdot k_{R2})
            \right]
     }
=1
\label{Phase_comm_closed}
\nonumber\\
\end{eqnarray}
The quantities entering the previous equation can be
trivially evaluated, in particular for any compact direction we have
\begin{equation}
\label{k_products}
2\alpha'\, (k_{L1}^a k_{R2}^b-k_{R1}^a k_{L2}^b)=
 w_1^a  n_2^b \frac{R^a}{R^b} - w_2^b  n_1^a \frac{R^b}{R^a}
\,\, ,\,\,
2\alpha'\, (k_{L1}^a k_{L2}^b-k_{R1}^a k_{R2}^b)=
 w_1^a  n_2^b \frac{R^a}{R^b} + w_2^b  n_1^a \frac{R^b}{R^a}
\end{equation}
Considering the cases where the momentum and
winding are different from zero only in a given compact direction $a$
we deduce immediately that we
have to impose the following restriction
on the coefficients entering the cocycles 
definition (\ref{c_closed}) 
\begin{equation}
\label{const_closed}
(\cc-\dd^T)_{aa}= 4{ N_{a}} +2
\,\,\,\,
{N_{a}}\in \interi,
\end{equation}
and
\begin{equation}
\bb_{a a},\ee_{a a}\in R ~.
\end{equation}
If we consider the cases where only two compact directions have
$k_{L/R}\ne 0$ and we choose their radii not to be equal 
we deduce that the off-diagonal elements 
must be identically zero:
\begin{eqnarray}
(\cc-\dd^T)_{a b}&=&0 ~~~~ (a\ne b) 
\nonumber\\
\bb_{a b}&=& \ee_{a b}=0 ~.
\end{eqnarray}

A further constraint comes from hermitian property of the vertex: 
the hermitian of a vertex is given by the following
expression 
\begin{equation}
\label{Herm_closed}
\left[{\cal W}^{(c)}_{\beta_L,\beta_R}(z,\bar z; k)\right]^\dagger
=
e^{i\Phi_{(c)}(k,k)}
\frac{1}{|z|^2}{\cal W}^{(c)}_{\beta_L,\beta_R}(\frac{1}{\bar z},\frac{1}{z}; -k)
\end{equation}
from which we find 
\begin{eqnarray}
e^{i\Phi_{(c)}(k,k)}
&=&
e^{-i\pi\, \alpha'\left[
 \bb\, k_{L} \cdot k_{L}  
+(\cc +\dd^T) \, k_{R} \cdot k_{L}  
+\ee \, k_{R} \cdot k_{R} 
\right]
}
\nonumber\\
&=&
e^{-i\pi\, \alpha'\left[
 \bb_{a a}\, k_{L a}^2  
+(\cc +\dd^T) \, k_{R a}  k_{L a}  
+\ee_{a a} \, k_{R a}^2 
\right]
}
\nonumber\\
&=&
e^{-i\pi\, \alpha'
[ \bb +\ee +  (\cc +\dd) ]_{a a}  \left(\frac{n_a}{R}\right)^2
}  
~e^{-i\pi\, \alpha'
[ \bb +\ee -  (\cc +\dd) ]_{a a}  \left(\frac{w_a R}{\alpha'}\right)^2
}  
~e^{-i\pi\,
[ \bb -\ee ]_{a a}  n_a w_a
}
\nonumber\\  
&=&1
\end{eqnarray}
and finally\footnote{
We could  also introduce a non operatorial cocycle also for the closed
vertices and then avoid these conclusions but this is an avoidable complication.}
\begin{equation}
\cc_{a a}= - \dd_{a a} = 1+ 2 N_a
~~~~~
\bb_{a a}=-\ee_{a a}= 2 M_a
\label{calC-calB}
\end{equation}
where $M_a$ are arbitrary integers.
In the text we will use
\begin{equation}
N_a=0
~~~~
M_a=0
\label{choice-closed-coc}
\end{equation}

Further constraints come from unitarity which requires that 
any three points amplitude ${\cal A}(1,2,3)$ be connected with the amplitude
${\cal A}(-1,-2,-3)$ for the CPT conjugate particles ($(-1)$ means
the CPT conjugate state of $1$) by
\begin{equation}
{\cal A}(1,2,3)=\left({\cal A}(-1,-2,-3)\right)^*
\end{equation}
This equation can be specialized to the case of three tachyons with
momenta $k_i=(k^\alpha_i,(k^a_{L i},k^a_{R i}))$ ($i=1,2,3$) 
and results in 
\begin{equation}
\label{3T}
{\cal A}(T_1,T_2,T_3)\propto
e^{i\sum_{1\le i\le j\le 3}\Phi_{(c)}(k_i,k_j)}
\delta^d(\sum k^\alpha_i) \delta_{\sum k_{L i}} \delta_{\sum k_{R i}}
\end{equation}
If we consider sequentially configurations with only one winding, two
windings different from zero, analogously for momenta and finally
configurations with mixed
momenta and windings   and 
we require a smooth behavior of the cocycles  
in the large and small $R_a$ limits we get the results given in the
main text.

%%%%%%%%%%%%%%%%%%%%%%%%%%%%%%%%%%%%%%%%%%%%%%%%%%%%%%%%%%%%%%%%%%%%%%
%%%%%%%%%%%%%%%%%%%%%%%%%%%%%%%%%%%%%%%%%%%%%%%%%%%%%%%%%%%%%%%%%%%%%%

\subsection{Open string computation.}
\label{details_Ostr}
In this appendix we give the details on how to obtain eq.s (\ref{VV-0},\ref{V-0})
and eq. (\ref{V-pi}).
To this purpose we write the generic closed string emission vertex in open
string formalism with the cocycles as
\begin{eqnarray}
\label{W_open}
{\cal W}_{\beta_L,\beta_R}(z,\bar z; k)
&=&
c(k_L,k_R,p)
{ V}_{\beta_L}(X_L(z);k_L)
{}~{V}_{\beta_R}(X_R(\bar z); k_R)
\end{eqnarray}
with
\begin{eqnarray}
c(k_L,k_R,p)
&=&
c_{nop}(k_L,k_R) c_{op}(k_L,k_R;p)
\nonumber\\
 c_{op}(k_L,k_R;p)
&=&
e^{i\pi\,\alpha' (u k_L+v k_R)\cdot p}
\\
c_{nop}(k_L,k_R)
&=&
e^{i\pi\alpha'\, (
\eta k_L \cdot k_L +\gamma k_L \cdot k_R +\epsilon k_R \cdot k_R
)}
,~~~~
\eta^T=\eta,~ \gamma^T=\gamma
\label{c_open}
\end{eqnarray}
where $c_{op}$ ($c_{nop}$) is the (non) operatorial part of the cocycle  
and $k_L=k_R=\frac{k}{2}$ in non compact direction.
Our aim is now to determine the (matrix) coefficients
$u,v,\eta,\gamma,\epsilon$ 
in order to reproduce the commutativity, the ``OPE'' coefficients in eq.
(\ref{OPE_closed}) or eq. (\ref{OPE_phase_closed_app0}) 
and the behavior of closed string vertices under
hermitian conjugation (\ref{Herm_closed}). 
It is worth stressing that the previous expression for the cocycles
(\ref{c_open}) contains a contribution from all directions, even the 
directions with $DD$ boundary condition since the
$\sigma$ coefficient in the string modes expansion is
an operator and hence commuting the cocycles with $X_{L/R}$ yields a
non trivial phase. 
If compute the phase  which we obtain normal ordering the product of
two vertices as done in (\ref{OPE_closed})
and we remember that $X_R=S X_R^{NN}$ (where $X_R^{NN}$ is the right
moving part with NN boundary conditions)
we get
\begin{equation}
\label{OPE_phase_open}
e^{i\Phi_{(o)}(k_1,k_2)}
=
\frac{c_{nop}(k_1) c_{nop}(k_2)}{c_{nop}(k_1+k_2)}
 e^{-i\pi\alpha'(u k_{L 2}+v k_{R 2})
      \cdot(k_{L 1}+S k_{R 1} )}
e^{  -i2\pi\alpha' Sk_{R 1}\cdot k_{L 2} 
}
\end{equation}
In this expression
 the first factor of rhs is due to non operatorial cocycles, the
second to the reordering of the operatorial ones and the last to the
reordering of $V_R(k_{R1}) V_L(k_{L2})$.
We require then that the closed string phase
(\ref{OPE_phase_closed_app}) in closed string formalism and
the corresponding phase
(\ref{OPE_phase_open}) in open string formalism be equal, 
i.e. 
\begin{equation}
\label{PhiO=PhiC}
e^{i\Phi_{(o)}(k_1,k_2)}=e^{i\Phi_{(c)}(k_1,k_2)}
\end{equation}  
which is the necessary condition for the correct factorization of
mixed open/closed string amplitudes in the closed channel. 
Given this basic constraint (\ref{PhiO=PhiC})
we can check that the further constraints  which arise from requiring
the commutativity of two closed string vertices and
the proper behavior under hermitian conjugation are identically
satisfied.
In particular the phase which arises when computing the hermitian
of a vertex is $\Phi_{(o)}(k,k)$
analogously to what we have found with the closed
string formalism (\ref{Herm_closed}).

The constraint (\ref{PhiO=PhiC}) can be explicitly rewritten as
\begin{eqnarray*}
&&
e^{ -i \pi\, \alpha'\left[
\left( \bb\, k_{L2} +\cc \, k_{R2}\right) \cdot k_{L1} 
+\left( \dd\, k_{L2} +\ee \, k_{R2}\right) \cdot k_{R1} 
\right] }
=
\\
&&
e^{ -i \pi\, \alpha'\left[
2\eta k_{L1} \cdot k_{L2} 
+2\epsilon k_{R 1} \cdot k_{R 2}
+\gamma k_{L1} \cdot k_{R2} 
+\gamma k_{L2} \cdot k_{R1} 
+ (u k_{L 2}+v k_{R 2})      \cdot(k_{L 1}+S k_{R 1})
+2 Sk_{R 1}\cdot k_{L 2}
\right]}
\end{eqnarray*}
By considering different combinations of windings and momenta as
done for the closed string cocycles after eq. (\ref{Phase_comm_closed})
and using eq.s (\ref{k_products}) 
we derive the following constraints:
\begin{equation}
\left(2\eta+(1+S)u+2S\right)_{\alpha\beta}=0,
\,\,\,\,
\gamma_{\alpha\beta}=\epsilon_{\alpha\beta}=v_{\alpha\beta}=0
\end{equation}
(up to a ``gauge'' choice which allows to set $\gamma=\epsilon=v=0$) 
for matrix indexes in the non compact directions and 
\begin{eqnarray}
||\left(2\eta+u -\bb \right)_{a b}||=
||\left(\ee-2\epsilon-Sv \right)_{a b}||
=2\,diag(\widehat M_{a}+ 2 \widehat D_a)
\end{eqnarray}
for  $k_{L 1 a} k_{L 2 b}$ and $-k_{R 1 a} k_{R 2 b}$ coefficients
in compact directions and
\begin{eqnarray}
||\left(\gamma^T+v-\cc \right)_{a b}||=
||-\left(-\dd+\gamma+Su+2S \right)_{a b}|| =2\,diag(\widehat M_{a})
\nonumber \\
\,\label{const_open}
\end{eqnarray}
for  $k_{L 1 a} k_{R 2 b}$ and $-k_{R 1 a} k_{L 2 b}$ coefficients.
The $\widehat M_a$  are arbitrary integers  because of eq.s (\ref{k_products}).
Consistency of the two last equations (\ref{const_open}) 
for $\gamma$ implies that
\begin{eqnarray}
v -(S u)^T &=& 2 +2 S + 4 \hat M + 4N
\label{cons-cond-gamma}
\end{eqnarray}

Next we want to exam the open string/closed string vertices commutativity.
To this purpose we write the open string emission vertex from the
$\sigma=0$ boundary as
\begin{equation}
{\cal V}_{\alpha}(x; k)
=
e^{i\pi\alpha'\, \tilde u k \cdot p}
{ V}_{\alpha}(X_L(x)\YZp;k)
\,~~~~x>0
\label{V-0-app}
\end{equation}
where $\tilde u$ is an unknown  matrix.
The naive way of writing this commutativity condition (which is wrong!) 
\begin{eqnarray*}
\left[
{\cal W}_{\beta_L,\beta_R}(X_L(z),X_R(\bar z))
\, 
{\cal V}_{\alpha}(x; k)
\right]_{an. con.}
&=&
{\cal V}_{\alpha}(x; k)
\,
{\cal W}_{\beta_L,\beta_R}(X_L(z),X_R(\bar z))
\end{eqnarray*}
implies that
\begin{equation}
\label{open_closed_comm}
e^{i\pi\alpha'\left[
(u k_L+v k_R )\cdot k
+2 k\cdot(k_L-Sk_R)
-\tilde u k\cdot(k_L+Sk_R)
\right]
}=1
\end{equation}
which can be satisfied when
\begin{eqnarray}
\label{const_open-closed_nc}
\left(-u+\tilde u^T(1+S)-2(1-S)\right)_{\alpha\beta} &=&0
\\
\label{const_open-closed}
\left(u-\tilde u^T +2\right)_{a b} =
\left(v-(\tilde u^T +2)S\right)_{a b} &=&0
\end{eqnarray}
respectively for matrix indices in non compact and compact directions.
In particular the equations in the last line arise
since eq. (\ref{open_closed_comm}) must be valid also when Wilson
lines are turned on, i.e. when $ k =\frac{n+\theta}{R}$.

In a similar way we get further constraints requiring that
the generic vertex for the open string emission from the $\sigma=\pi$ border
\begin{equation}
{\cal V}_{\alpha}(-x; k)
=
e^{i\pi\alpha'\, \tilde u_\pi k \cdot p}
{ V}_{\alpha}(X_L(-x)\YZp;k)
\end{equation}
commute with both closed string and $\sigma=0$ open string vertices.

%%%%%%%%%%%%%%%%%%%%%%%%%%%%%%%%%%%%%%%%%%%%%%%%%%%%%%%%%%%%%%%%%%%%%%

The coefficient $\tilde u_\pi$ can be fixed by
requiring the commutativity of $\sigma=\pi$ open string vertices
the with $\sigma=0$ open string vertices;
it turns easily out that
\begin{equation}
\tilde u_\pi=\tilde u - 2
\end{equation}
This in turn implies that the commutativity of an
open string emission vertex from $\sigma=\pi$ with a closed string
vertex does not really yield new constraints but some consistency
conditions which read 
\begin{eqnarray}
\label{const_open_pi-closed_nc}
\left(-u+\tilde u^T(1+S))\right)_{\alpha\beta} &=&0
\\
\label{const_open_pi-closed}
\left(u-\tilde u^T \right)_{a b} =
\left(v-\tilde u^T S\right)_{a b} &=&0
\end{eqnarray}

Unfortunately or better correctly  the sets of constraints
(\ref{const_open-closed_nc}-\ref{const_open-closed}) and  
(\ref{const_open_pi-closed_nc}-\ref{const_open_pi-closed}) 
are {\sl inconsistent}.
This would seem to be a disaster but luckily it is not so 
since the proper way of formulation open/closed string commutativity reads, as 
explained in the main text, is to take into account the $y_0$ shift so that
\begin{eqnarray*}
\left[
{\cal W}_{\beta_L,\beta_R}(X_L^{out}(z),X_R^{out}(\bar z))
\, 
{\cal V}_{\alpha}(x; k)
\right]_{an. con.}
&=&
{\cal V}_{\alpha}(x; k)
\,
{\cal W}_{\beta_L,\beta_R}(X_L^{in}(z),X_R^{in}(\bar z))
\nonumber \\
&=&
{\cal V}_{\alpha}(x; k)
\,
{\cal W}_{\beta_L,\beta_R}(X_L^{out}(z),X_R^{out}(\bar z))
e^{2\pi \alpha' i\,k\cdot(k_L-Sk_R)}
\nonumber\\
\end{eqnarray*}
As a consequence of this proper commutativity condition 
all the constraints from open/closed string vertices
commutativity are now consistent and turn out to be eq.s
(\ref{const_open_pi-closed}) because of the extra phase contribution
from the proper commutativity condition.

From hermiticity conjugation of ${\cal V}_\alpha$ vertices we get 
\begin{equation}
\tilde u+\tilde u^T =0
\end{equation}
which gives using eq.s (\ref{const_open_pi-closed_nc}),
(\ref{const_open_pi-closed}) and the consistency condition (\ref{cons-cond-gamma})
\begin{equation}
\tilde u = u = v =0
~~~~~~
2 \widehat M + 2 N + {1+S}=0
\end{equation}
Finally using eq.s (\ref{calC-calB}) we get
\begin{eqnarray}
\gamma
&=& 
(\cc+ 2 \widehat M)^T
=
(1+2 N + 2 \widehat M)^T
=
-S
\nonumber\\
\eta&=&-\epsilon
=\frac{1}{2} \bb + \widehat M + 2 \widehat D
= - \frac{1}{2} \ee + \widehat M + 2 \widehat D
= M + \widehat M + 2 \widehat D
\end{eqnarray}
With our choice (\ref{choice-closed-coc}) and choosing $\widehat
M=\widehat D=0$ we finally find
\begin{equation}
\gamma=-S,
~~~~
\eta=-\epsilon=0
\end{equation}

It is interesting to notice that the hermiticity of the open string
vertices
${\cal V}_\alpha(-x)$ does not yield any further constraint since the
phase $e^{i 2\pi \alpha' k^2}$ which is
obtained reordering the cocycle $e^{i\pi\alpha'\, \tilde u_\pi k \cdot p}$
after taking the hermitian conjugate is precisely the one needed to
express the hermitian  as a function of $\frac{1}{(-x)^*}
=\frac{1}{(  e^{i \pi} |x|)^*}= \frac{e^{i \pi}}{|x| }$; explicitly we
get
\begin{equation}
{\cal V}_\alpha(-x;k)^\dagger 
= 
e^{i 2\pi \alpha' k^2} \left(\frac{1}{x} \right)^{ 2\alpha' k^2} 
{\cal V}_\alpha(\frac{1}{(-x)^*};-k) 
= \left(\frac{e ^{i \pi}}{|x|} \right)^{ 2\alpha' k^2}
{\cal V}_\alpha(\frac{1}{(-x)^*};-k) 
\end{equation}

%%%%%%%%%%%%%%%%%%%%%%%%%%%%%%%%%%%%%%%%%%%%%%%%%%%%%%%%%%%%%%%%%%%%%%
%%%%%%%%%%%%%%%%%%%%%%%%%%%%%%%%%%%%%%%%%%%%%%%%%%%%%%%%%%%%%%%%%%%%%%
%\section{Details on amplitudes factorizations.}

\section{Changing variables in the open string amplitude.}
\label{App:ChangeVar}
In this appendix we would like to discuss some subtleties in performing 
the $SL(2,\complessi)$ change of variable 
$w={z+{\rm i} \over z-{\rm i}}~$, which is not a symmetry, 
on the correlator (\ref{corr-z}).
Naively one would say that the answer is
\begin{equation}
\label{corr-w}
\prod\theta(\phi_{i}- \phi_{i+1})
\langle 0 |~{\rm T}\left(
 \prod_{i=1}^{N_o} { V}_{\alpha_i}(e^{i\phi_i};p_i)de^{i\phi_i}
\,\prod_{j=1}^{N_c} 
  {\cal W}_{\beta_{L j},\beta_{R j}}(w_j, \frac{1}{\bar w_j}; k_j)
 \frac{d^2w_j}{\bar w^2_j}
\right)
| 0\rangle~~
\end{equation}
where we now use open string vertices with the trivial cocycle because
there is not anymore any difference between emission from real
positive axis and real negative one.

If we look more carefully we realize that the two correlators are related
in a non trivial way by various analytic continuations:
\begin{itemize}
\item in the $SL(2,\complessi)$ parameters entering the operator 
realizing the wanted $SL(2,\complessi)$ transformation 
%(see the appendix B for more details)
;
\item  in the order of the operators since it can happen that a
$SL(2,\complessi)$ transformation changes the radial ordering, i.e. it
does not preserve the absolute of $|z|$; 
\item in the log which enters the string expansion  since 
$w$ and $\bar w$ can move independently on different sheets 
because the phases of $w$ and $\bar w$ can exceed the range $]-\pi,\pi]$ and 
the phase of $\bar w$ can also be  not the opposite of $w$.
\end{itemize}
Moreover there is a further overall phase ambiguity due to the
possible ordering choices of open string vertices whose Euclidean
times are now the same, of which
eq. (\ref{corr-w}) is a particular one.

Since it is very difficult to keep track  of these
analytic continuations, we are therefore led to use an indirect way 
to compute the phase which arises performing the transformation (\ref{transf}) 
 on eq. (\ref{corr-z})%, or more generally to the final result
:
we compare the explicit expression for (\ref{corr-z}) obtained
using the Reggeon vertex formalism on which we perform the
change of variables (\ref{transf}) with the corresponding explicit
result of the amplitude ($\phi_{N_o+1}=\pi$)
\begin{eqnarray}
\label{corr-wc}
A(N_o, N_c)&=&
e^{i\alpha_0}
\langle 0 |~{\rm T}\Big(
 \prod_{i=1}^{N_o} \EYZi %e^{i p_{i} y_0} 
{ V}_{\alpha_i}(e^{i\phi_i};p_i)de^{i\phi_i}
\,\prod_{j=1}^{N_c} 
  c_L(k_{L j},k_{R j};p)\,e^{i k_{L j} y_0} 
  { V}_{\beta_{L j}}(w_j; k_{L j}) dw_j
\nonumber\\
&&
\,\prod_{l=1}^{N_c} 
  c_R(k_{L l},k_{R l};-Sp)\, e^{-i S k_{R j} y_0} 
 { V}_{\beta_{R l}}(\frac{1}{\bar w_l};
  k_{R l})
 \frac{d\bar w_l}{\bar w^2_l}
\Big)
| 0\rangle~~
\end{eqnarray}
where we have made explicit the $y_0$ dependence by redefining the
$X_{L,R}$ in such a way they are free of $y_0$ 
and we have introduced the cocycles $c_L$ and $c_R$\footnote{
These cocycles are the same as in closed string (\ref{cocycl-closed})
with the substitution
of the closed string operators $p_L, p_R$ with the open string operator $p$.
} to ensure that
the amplitude can be obtained by  the analytic continuation of 
whichever radial order
we choose or, said in different words, commuting 
${V}_{\beta_{L j}}$ and ${V}_{\beta_{L l}}$  produces a phase 
$$e^{i\pi\, \alpha'
            \left(
            (\cc k_{L l}\cdot k_{R j}-k_{R l}\cdot \cc k_{L j})
            +(\bb k_{L j}\cdot k_{L l}-\bb k_{L l}\cdot k_{L j})
            +2 sgn(arg(w_j)-arg(w_l)) k_{L j}\cdot k_{L l}
            \right)
     }
$$
which is canceled by the phase 
$$
e^{i\pi\, \alpha'
            \left(
             (\dd k_{L j}\cdot k_{R l}-k_{R j}\cdot \dd k_{L l})
            -(\ee k_{R l}\cdot k_{R j}-\ee k_{R j}\cdot k_{R l})
            -2 sgn(arg(w_j)-arg(w_l))k_{R j}\cdot k_{R l}
            \right)
     }
$$ 
we get while commuting  the corresponding right vertex
operators.
To derive the last equation we have made use of the fact that
$arg(\bar w)=-arg(w)$; as it was stressed before this relation could
not otherwise have been taken for granted if we had performed the
transformation by inserting the $SL(2,\complessi)$ operators.
It is important to stress that thanks to the cocycle the phase $\alpha_0$
is independent of the different specific cases which arise from the 
radial ordering.
%The phase $\alpha_0$ in eq. (\ref{corr-wc}) is obtained by direct comparison of
%the actual values of the amplitudes eq.s (\ref{corr-z}) and
%(\ref{corr-wc}) and it is
%\begin{equation}
%\alpha_0=
%\end{equation}

\subsection{Determining the phase $e^{i\alpha(p)}$. }
\label{app_deter_phase}
In order to compute the phase $e^{i\alpha(p)}$ we generalize the
computation in section \ref{1Tc1To} and we compare the $1$
closed tachyon- $N_o$ open tachyons amplitude at the tree level
computed in open and closed string formalism.
This is enough to fix the phase since any amplitude involving a boundary state 
can be factorized on 
a one point closed string with boundary amplitude
times a closed string amplitude 
and phases arise only from momenta contributions. 

The $1$ closed tachyon- $N_o$ open tachyons amplitude in open string 
formalism when all directions are compact reads 
\begin{eqnarray}
\label{1TcNTo}
{\cal A}(N_o T_o, 1 T_c ) & = &
{\cal C}_{0(p)}\,{\cal N}_{o(p)}^{N_o} 
\,{{\cal N}_c}
\int 
d x_{N_o}
~\prod_{i=2}^{N_o-1} \left[d x_i\;\theta(x_{i+1}- x_i)\right]
~(-2)
\nonumber\\ 
& & 
2^{2\alpha' k_L \cdot S k_R} e^{i y_0\cdot (k_L-S k_R)}
~\prod_{2\le i<j \le N_o} (x_i-x_j)^{2\alpha' k_i \cdot k_j}
\nonumber\\
&&
~\prod_{2\le j \le N_o} (x_j-i)^{2\alpha' k_j \cdot k_L} 
(x_j+i)^{2\alpha' k_j \cdot k_R}
\delta_{k_L+S k_R+ \sum_{j=1}^{N_o} k_j,0}
\nonumber\\
\end{eqnarray}
where we have gauge fixed the $SL(2,R)$ invariance with $z=i$ and
$x_1\rightarrow+\infty$. The contribution $(-2)$ is what is left 
of $d V_{a b c}$ and $2^{2\alpha' k_L \cdot S k_R}$ is the
closed string tachyon contribution after the gauge fixing.

Changing variables to the disk ones as described in the main text, i.e.
$ w={z+{\rm i} \over z-{\rm i}}$ so that $w(x)=e^{i\phi}$
and using the momentum conservation and mass shell conditions
 we get
\begin{eqnarray}
\label{1TcNTo-cyl}
{\cal A}(N_o T_o, 1 T_c ) & = &
{\cal C}_{0(p)}\,{\cal N}_{o(p)}^{N_o} 
\,{{\cal N}_c}
~i~(i)^{-2\alpha' k_L \cdot S k_R}
~\int_0^{\pi} 
~\prod_{j=2}^{N_o} \left[d e^{i \phi_j} \;\theta(\phi_{j+1}- \phi_j)\right]
\nonumber\\ 
& & 
2^{2\alpha' k_L \cdot S k_R} e^{i y_0\cdot (k_L-S k_R)}
~\prod_{2\le i<j \le N_o} (e^{i \phi_i} - e^{i \phi_j})^{2\alpha' k_i \cdot k_j}
~\prod_{2\le j \le N_o} (e^{i \phi_j})^{2\alpha' k_j \cdot k_L} 
\nonumber\\
&&
\delta_{k_L+S k_R+ \sum_{j=1}^{N_o} k_j,0}
\end{eqnarray}
where we have set $\phi_1=0$.
The corresponding amplitude computed using the boundary is given by
\begin{eqnarray}
\label{1TcNTo-bou}
{\cal A}(N_o T_o, 1 T_c ) & = &
%{\widehat \cc}_0
2\pi \frac{T_p}{2}\,{\cal N}_{o(p)}^{N_o} 
~i~e^{i \alpha(k) }
~\int_0^{\pi} 
~\prod_{j=2}^{N_o} \left[d e^{i \phi_j} \;\theta(\phi_{j+1}- \phi_j)\right]
\nonumber\\ 
& & 
2^{2\alpha' k_L \cdot S k_R} e^{i y_0\cdot (k_L-S k_R)}
~\prod_{2\le i<j \le N_o} (e^{i \phi_i} - e^{i \phi_j})^{2\alpha' k_i \cdot k_j}
~\prod_{2\le j \le N_o} (e^{i \phi_j})^{2\alpha' k_j \cdot k_L} 
\nonumber\\
&&
\delta_{k_L+S k_R+ \sum_{j=1}^{N_o} k_j,0}
\end{eqnarray}

Comparing these two expressions 
%and since any amplitude involving a boundary state can be factorized on a
%one point closed string amplitude
it follows that the generic phase can be written in an operatorial way
as
\begin{equation}
e^{i \alpha(p)}=e ^{-i \pi \alpha' p_L \cdot S p_R}
\end{equation}
where the operatorial momenta $p_L$ and $p_R$ must be identified in non compact
directions.

%%%%%%%%%%%%%%%%%%%%%%%%%%%%%%%%%%%%%%%%%%%%%%%%%%%%%%%%%%%%%%%%%%%%%%
%%%%%%%%%%%%%%%%%%%%%%%%%%%%%%%%%%%%%%%%%%%%%%%%%%%%%%%%%%%%%%%%%%%%%%
%%%%%%%%%%%%%%%%%%%%%%%%%%%%%%%%%%%%%%%%%%%%%%%%%%%%%%%%%%%%%%%%%%%%%%
%%%%%%%%%%%%%%%%%%%%%%%%%%%%%%%%%%%%%%%%%%%%%%%%%%%%%%%%%%%%%%%%%%%%%%

\end{document}